\documentclass[
reprint,
superscriptaddress,
amsmath,
amssymb,
aps,
prb,
longbibliography,
noeprint
]{revtex4-2}
\usepackage{graphicx}% Include figure files
\usepackage{dcolumn}% Align table columns on decimal point
\usepackage{bm}% bold math
\usepackage{needspace}% for proper figure enforcement

\usepackage{hyperref}
\usepackage{xcolor}
\hypersetup{
    colorlinks,
    linkcolor={red!50!black},
    citecolor={blue!50!black},
    urlcolor={blue!80!black}
}
\usepackage[capitalise]{cleveref}

\newenvironment{myfont}{\fontfamily{phv}\selectfont}{\par}
\makeatletter
\renewcommand\frontmatter@abstractwidth{\dimexpr\textwidth-0.5in\relax}
\makeatother

\newcommand{\um}{\ensuremath{ \mu \mathrm{m} }}

\newcommand{\js}{\ensuremath{ \bm{j}_\mathrm{s} }}
\newcommand{\Idc}{\ensuremath{ I_\mathrm{dc} }}
\newcommand{\phiH}{\ensuremath{ \phi_\mathrm{H} }}
\newcommand{\thetaH}{\ensuremath{ \theta_\mathrm{H} }}

\newcommand{\taust}{\ensuremath{ \bm{\tau}_\mathrm{st} }}
\newcommand{\tauH}{\ensuremath{ \bm{\tau}_\mathrm{H} }}
\newcommand{\taudamp}{\ensuremath{ \bm{\tau}_\mathrm{d} }}
\newcommand{\taua}{\ensuremath{ \bm{\tau}_\mathrm{a} }}

\newcommand{\dRac}{\ensuremath{ \delta R_\mathrm{ac} }}

\newcommand{\Ms}{\ensuremath{ M_\mathrm{s} }}
\newcommand{\Mx}{\ensuremath{ M_\mathrm{x} }}
\newcommand{\My}{\ensuremath{ M_\mathrm{y} }}
\newcommand{\Mz}{\ensuremath{ M_\mathrm{z} }}

\newcommand{\Mvec}{\ensuremath{ \bm{M} }}
\newcommand{\Hvec}{\ensuremath{ \bm{H} }}
\newcommand{\Ha}{\ensuremath{ \bm{H}_\mathrm{a} }}

\newcommand{\HDx}{\ensuremath{ H_\mathrm{D_x} }}
\newcommand{\HDy}{\ensuremath{ H_\mathrm{D_y} }}
\newcommand{\HDz}{\ensuremath{ H_\mathrm{D_z} }}
\newcommand{\HPMA}{\ensuremath{ H_\mathrm{PMA} }}

\newcommand{\Dx}{\ensuremath{ D_\mathrm{x} }}
\newcommand{\Dy}{\ensuremath{ D_\mathrm{y} }}
\newcommand{\Dz}{\ensuremath{ D_\mathrm{z} }}

\begin{document}

\preprint{APS/123-QED}

\title{\myfont Easy-plane spin Hall oscillator}

\author{\myfont Eric Arturo Montoya}
\email{eric.montoya@utah.edu}
\affiliation{\myfont Department of Physics and Astronomy, University of California, Irvine, California 92697, USA}
\affiliation{\myfont Department of Physics and Astronomy, University of Utah, Salt Lake City, Utah 84112, USA}

\author{\myfont Amanatullah Khan}
\affiliation{\myfont Department of Physics and Astronomy, University of California, Irvine, California 92697, USA}

\author{\myfont Christopher Safranski}
\affiliation{\myfont Department of Physics and Astronomy, University of California, Irvine, California 92697, USA}

\author{\myfont Andrew Smith}
\affiliation{\myfont Department of Physics and Astronomy, University of California, Irvine, California 92697, USA}

\author{\myfont Ilya N. Krivorotov}
\email{ilya.krivorotov@uci.edu}
\affiliation{\myfont Department of Physics and Astronomy, University of California, Irvine, California 92697, USA}

\date{\today}

\begin{myfont}

\begin{abstract}
\textbf{
Spin Hall oscillators (SHOs) based on bilayers of a ferromagnet (FM) and a non-magnetic heavy metal (HM) are electrically tunable nanoscale microwave signal generators.
Achieving high output power in SHOs requires driving large-amplitude magnetization dynamics by a direct spin Hall current. 
The maximum possible amplitude of such oscillations with the precession cone angle nearing 90$^{\circ}$ is predicted for FM layers with easy-plane magnetic anisotropy and spin Hall current polarization perpendicular to the easy plane. While many FMs exhibit natural easy-plane anisotropy in the FM film plane, the spin Hall current in a HM$|$FM bilayer is polarized in this plane and thus cannot drive large-amplitude magneto-dynamics. 
Here we present a new type of SHO engineered to have the easy-plane anisotropy oriented normal to the film plane, enabling large-amplitude easy-plane dynamics driven by spin Hall current. 
Our experiments and micromagnetic simulations demonstrate that the desired easy-plane anisotropy can be achieved by tuning the magnetic shape anisotropy and perpendicular magnetic anisotropy in a nanowire SHO, leading to a significant enhancement of the generated microwave power.
The easy-plane SHO experimentally demonstrated here is an ideal candidate for realization of a spintronic spiking neuron.
Our results provide a new approach to design of high-power SHOs for wireless communications, neuromorphic computing, and microwave assisted magnetic recording.
}
\end{abstract}
\maketitle
\end{myfont}

\begin{large}
\noindent \textbf{\myfont Introduction}
\end{large}

Manipulation of magnetization by spin-orbit torques (SOTs) forms the basis of several promising spintronic technologies such as spin-orbit torque memory (SOT-MRAM) \cite{mironPerpendicularSwitching2011, liuSpinTorqueSwitching2012, baumgartnerSpatiallyTimeresolved2017, satoTwoterminalSpin2018, finocchioSpinOrbit2020, zhengFieldfreeSpinorbit2021}, spin-orbit torque oscillators \cite{liuMagneticOscillations2012, safranskiSpinOrbit2019, haidarSingleLayer2019}, neuromorphic computing devices \cite{senguptaSpinorbitTorque2015, tsunegiPhysicalReservoir2019, grollierNeuromorphicSpintronics2020, hassanLowBarrierMagnet2019, zahedinejadMemristiveControl2022}, and SOT-based magnonic logic \cite{demidovSpinOrbittorque2020, chumakAdvancesMagnetics2022short}.
Additionally, spin-orbit torque oscillators serve as a test bed for fundamental studies of strongly nonlinear magnetization dynamics in nanoscale ferromagnets \cite{barsukovGiantNonlinear2019a}. 

\begin{figure}[tbph]
\centering
 \includegraphics[width= 0.9\columnwidth]{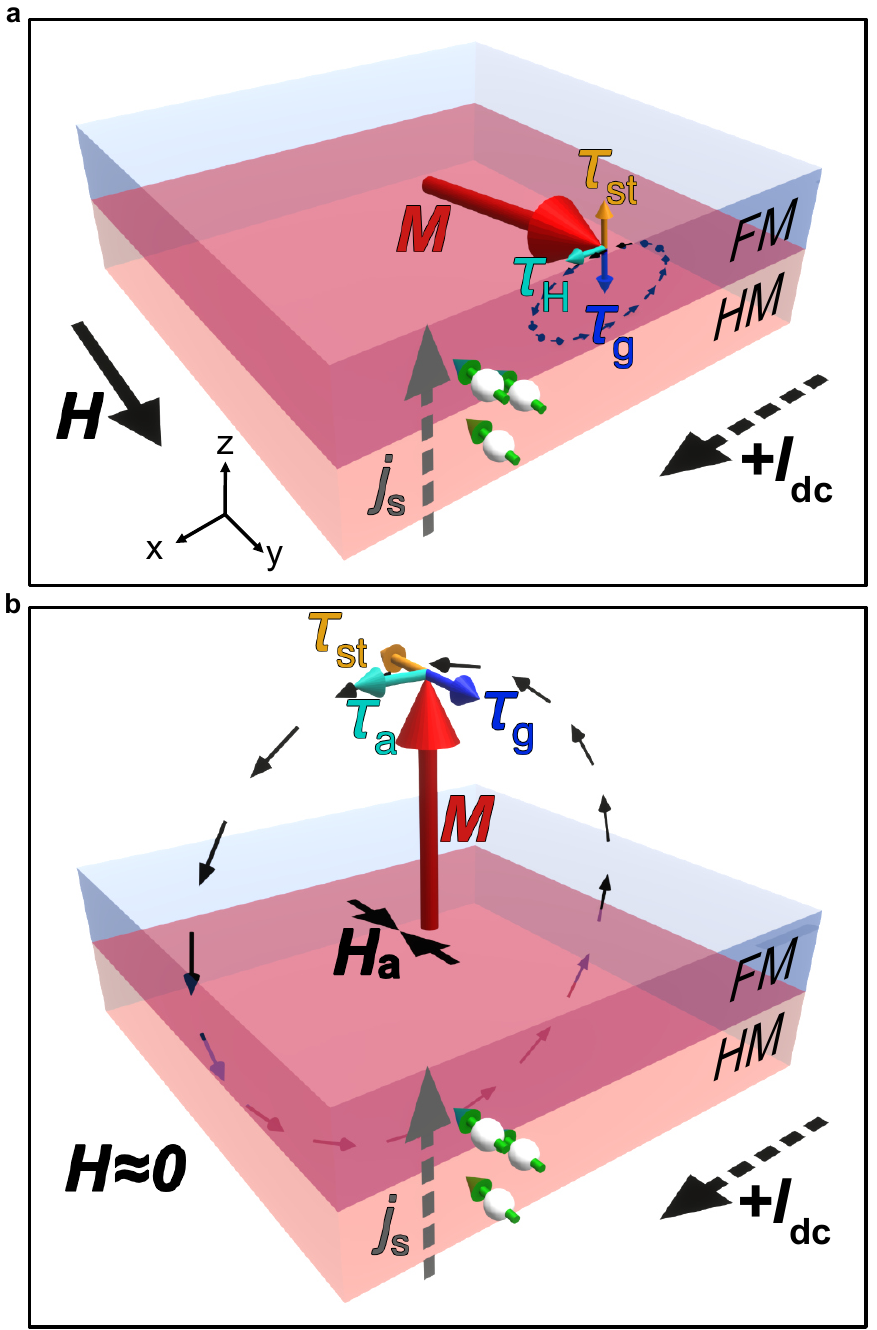}
 \caption{\textbf{Spin Hall oscillator dynamics.} \textbf{(a)} Persistent magnetization dynamics in a conventional spin Hall oscillator. 
 The spin Hall current $\js$ (gray dashed arrow) from heavy metal (HM) applies spin Hall torque $\taust$ (orange arrow) that compensates the Gilbert damping torque $\taudamp$ (dark blue arrow) and drives persistent precession (black dashed arrows) of the FM magnetization $\bm{M}$ (red arrow).
 The spin current is polarized in the plane of the FM film (green arrows) and an external field $\bm{H}$ is applied to define the precession axis (large black arrow), leading to small-angle precession of magnetization due to the effective field torque $\tauH$ (turquoise arrow). 
 \textbf{(b)} Easy-plane spin torque oscillator dynamics. In this geometry, easy-plane magnetic anisotropy is perpendicular to the FM layer plane and spin Hall current is polarized perpendicular to the easy plane. 
 Spin Hall torque $\taust$ pulls $\bm{M}$ out of the easy plane and the anisotropy torque $\taua$ (turquoise arrow) drives large-amplitude magnetization precession around the anisotrpy field $\Ha$ that is perpendicular to the easy plane.
\label{fig:SHO_types}}
 \end{figure}

The simplest type of spin-orbit torque oscillator is the spin Hall oscillator (SHO) \cite{demidovMagneticNanooscillator2012,duanNanowireSpin2014,giordanoSpinHallNanooscillator2014a, smithDimensionalCrossover2020, zhangSpintorqueOscillation2020, hacheBipolarSpin2020}. SHO is based on a bilayer of a ferromagnet (FM) and a non-magnetic heavy metal (HM), as illustrated in \cref{fig:SHO_types}(a). 
 A direct electric charge current in the HM layer flowing along the $x$-axis gives rise to a pure spin current density $\js$ along the $z$-axis (grey dashed arrow) with its magnetic  polarization parallel the $y$-axis (green arrows). 
 Interaction of $\js$ with the FM magnetization $\Mvec$ gives rise to spin Hall torque $\taust\sim\js$ opposing the Gilbert damping torque $\taudamp$. We use the term polarization to refer to direction of electron magnetic moment of spin current.

When $\js$ exceeds a critical value proportional to the FM Gilbert damping parameter $\alpha$, $\taust$ overcomes the damping $\taudamp$ and excites persistent auto-oscillatory magnetization precession (shown by small black arrows) around the equilibrium direction of $\Mvec$. 
The lowest critical current is observed for $\Mvec$ magnetized in the $-y$-direction by an applied field $\Hvec$. 

For easy-plane magnetic anisotropy coinciding with the FM layer plane, the precession trajectory is elliptical as shown in \cref{fig:SHO_types}(a).
The precession frequency increases with increasing $\Hvec$ and the anisotropy field $\Ha$, and decreases with increasing precession amplitude due to the negative nonlinear frequency shift in this geometry \cite{slavinNonlinearAutoOscillator2009}. 
The amplitude of  precession first increases with increasing $\js$ but then saturates at precession cone angles typically not exceeding 20$^{\circ}$ due to the nonlinear damping mechanism \cite{divinskiyControlledNonlinear2019, smithDimensionalCrossover2020, leeOriginNonlinear2022a}. 
The current-driven auto-oscillations of $\Mvec$ generate microwave voltage due to the FM magnetoresistance. 
The frequency and amplitude of this microwave voltage depend on $\js$ and thus SHOs are electrically tunable microwave signal generators of nanoscale dimensions \cite{slavinNonlinearAutoOscillator2009}.

SHO generators of microwave signals with high output power and low phase noise are desirable for applications \cite{senguptaSpinorbitTorque2015,gotoUncooledSubGHz2021}.
Large-amplitude persistent magnetization dynamics with the precession cone angle of nearly 90$^{\circ}$ has been predicted for easy-plane FMs upon injection of spin current polarized normal to the easy plane \cite{kentSpintransferinducedPrecessional2004, wangDynamicsThinfilm2006, rowlandsMagnetizationDynamics2012}. 
Such type of an easy-plane SHO (EP-SHO) with the magnetically easy $xz$-plane is shown in \cref{fig:SHO_types}(b).
In contrast to the conventional SHO in \cref{fig:SHO_types}(a), large-amplitude precessional dynamics in EP-SHOs is excited  immediately above the critical current \cite{kentSpintransferinducedPrecessional2004}. 

The EP-SHO magnetization is tilted out of the easy-plane by the spin Hall torque $\taust$ whereupon it precesses with large amplitude around the anisotropy field $\Ha$ parallel to the $y$-axis. \cite{firastrauModelingPerpendicular2008}. 
The critical current for these dynamics is defined by a smaller magnetic anisotropy present within the dominant easy-plane anisotropy rather than by the FM Gilbert damping \cite{kentSpintransferinducedPrecessional2004, wangDynamicsThinfilm2006, rowlandsMagnetizationDynamics2012}. The EP-SHO can operate in zero external magnetic field, which is desired for many applications. The EP-SHO system is especially attractive for realization of a magnetic spiking neuron as has been proposed in several recent theoretical publications \cite{khymynUltrafastArtificial2018,matsumotoChaosRelaxation2019, markovicEasyplaneSpin2022}. The EP-SHO is predicted to generate a large-amplitude sub-nanosecond output voltage pulse in response to an input current pulse exceeding a threshold value. Null output is expected for sub-threshold input pulses. This is the characteristic behavior of a spiking neuron.

Large-amplitude easy-plane persistent dynamics have been theoretically studied in spin-transfer-torque nanopillar devices \cite{firastrauModelingPerpendicular2008, rowlandsMagnetizationDynamics2012}, but have yet to be explored in SOT devices, such as SHOs.  
Here we report experimental realization of a nanowire EP-SHO based on a Pt$|$FM bilayer, where FM is a Co$|$Ni superlattice \cite{manginCurrentinducedMagnetization2006,rippardSpintransferDynamics2010,mohseniHighFrequency2011,mohseniSpinTorque2013,choiVoltagedrivenGigahertz2022,maciaStableMagnetic2014}. 
The EP-SHO dynamics is achieved via tuning the Co$|$Ni perpendicular magnetic anisotropy (PMA) and the magnetic shape anisotropy of the nanowire to manufacture an easy plane defined by the wire axis and the film normal ($xz$-plane) as shown in \cref{fig:Schematic_AMR}(a).
We present measurements and micromagnetic simulations demonstrating that the microwave power generated by the SHO is maximized when the magnetic easy-plane energy landscape is realized. 

Our results are relevant for engineering of scalable, high-power SHOs for wireless communications \cite{leePowerEfficientSpinTorque2019}, neuromorphic computing \cite{torrejonNeuromorphicComputing2017,grollierNeuromorphicSpintronics2020,flebusNonHermitianTopology2020}, room-temperature radio frequency bolometers \cite{gotoUncooledSubGHz2021}, and microwave assisted magnetic recording \cite{okamotoMicrowaveAssisted2015}. 

\textbf{\begin{figure*}[tbph]
\centering
 \includegraphics[width= 1.0\textwidth]{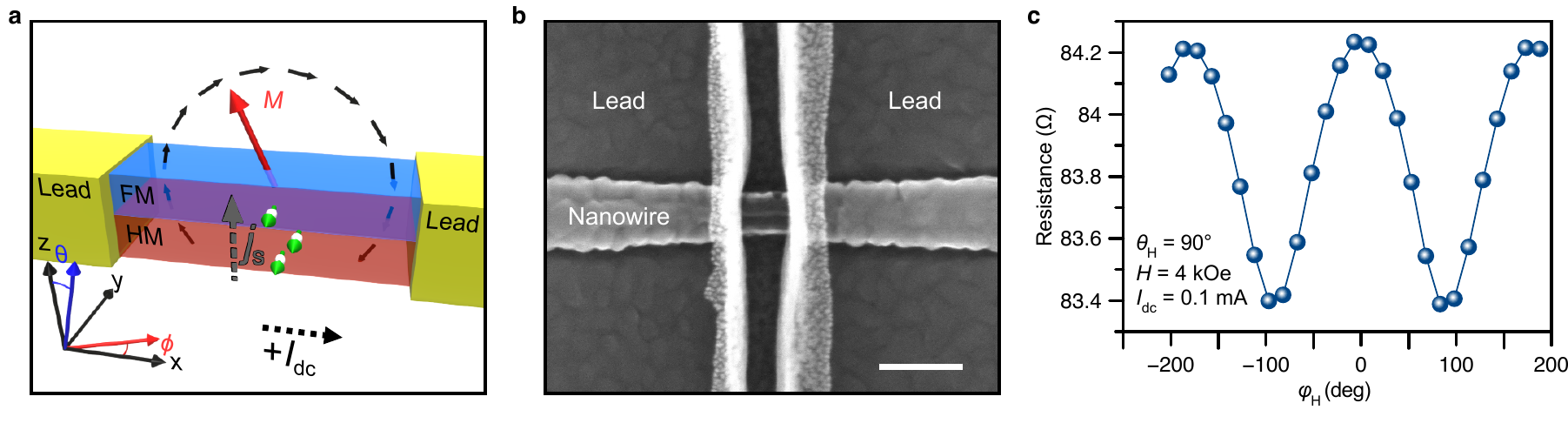}
 \caption{\textbf{Device schematics and magnetoresistance.} \textbf{(a)} Schematic of easy-plane spin Hall oscillator (EP-SHO) based on a heavy metal (HM=Pt) and a ferromagnetic metal (FM=Co$|$Ni superlattice) bilayer nanowire. 
 A positive direct charge current in the HM layer $+\Idc$ (black dashed arrow) generates a spin Hall current $\js$ (gray dashed arrow) flowing in the $z$-direction with its polarization in the $-y$-direction (green arrows). Spin current $\js$ impinging on the FM applies spin Hall torque $\taust$ to magnetization $\bm{M}$ and pulls it out of the easy $xz$-plane. The magnetization then precesses about the easy-plane anisotropy field $\Ha||y$ as indicated by black arrows.
 \textbf{(b)} Scanning electron micrograph of an EP-SHO. The scale bar is 100 nm. 
 \textbf{(c)} Resistance of the EP-SHO device in \textbf{(b)} as a function of a $4$\,kOe magnetic field direction in the $xy$-plane measured at $T=4.2$\,K.
\label{fig:Schematic_AMR}}
 \end{figure*}}

\textbf{\begin{figure*}[htbp]
\center
 \includegraphics[width= 1.0\textwidth]{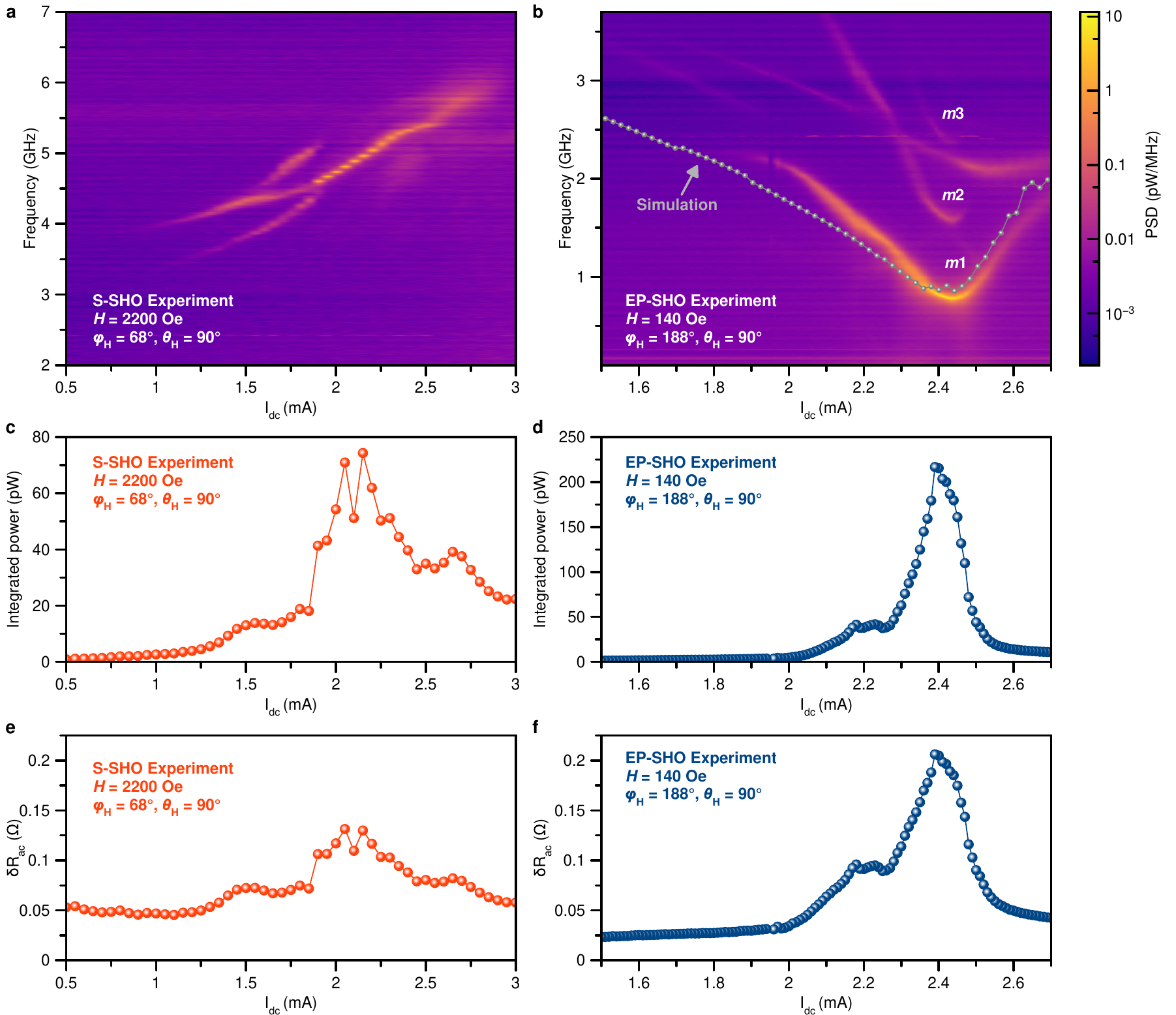}
 \caption{\textbf{Microwave signal emission.} Power spectral density (PSD) of microwave signal generated for the (\textbf{a}) high-field S-SHO configuration and (\textbf{b}) low-field EP-SHO configuration. Integrated power for the (\textbf{c}) S-SHO and (\textbf{d}) EP-SHO. 
 Amplitude of resistance oscillations for the (\textbf{e}) S-SHO and (\textbf{f}) EP-SHO. 
\label{fig:Emission_experiment}}
 \end{figure*}}

\Needspace{1\baselineskip} %To make sure it actually shows up after the figure
\begin{large}
\noindent \textbf{\myfont Results}
\end{large}

\noindent \textbf{\myfont Device geometry and magnetoresistance.}
\Cref{fig:Schematic_AMR}(a) shows a schematic of the nanowire easy-plane spin Hall oscillator device along with the coordinate system used in this article. 
An applied direct electric current flowing in HM Pt along the length of the nanowire ($x$-direction) leads to a transverse spin Hall current \cite{sagastaUnveilingMechanisms2018} flowing in the $z$-direction that is polarized in the $-y$-direction. 
When injected into the FM, the spin Hall current applies spin Hall torque to the FM magnetization \cite{bergerDeterminationSpin2018} and drives auto-oscillatory magnetization dynamics \cite{liuMagneticOscillations2012}. 

The EP-SHO nanowires studied here were patterned from substrate$||$seed$|$HM$|$FM$|$cap films deposited by magnetron sputtering. 
We used Pt(7 nm) for the HM layer and a $\left[\mathrm{Co}(0.98\,\mathrm{nm})|\mathrm{Ni}(1.46\,\mathrm{nm}) \right]_{2}|\mathrm{Co}(0.98\,\mathrm{nm})$ superlattice as the FM layer. 
The Co$|$Ni superlattice was selected for its large anisotropic magnetoresistance (AMR) and tunability of PMA via the Co and Ni layer thicknesses \cite{aroraOriginPerpendicular2017}.  Highly resistive Ta is employed for the seed (3 nm) and capping (4 nm) layers \cite{montoyaSpinTransport2016}. Electron beam lithography and Ar$^+$ ion milling were used to define 50 nm wide, 40 $\um$ long nanowires from the film stack. 
Ta(5 nm)$|$Au(40 nm)$|$Ta(5 nm) electric leads were attached to the nanowire with the inter-lead gap varying from 50 to 450 nm.
The spacing between the leads defines the active region of the nanowire where current density exceeding the critical value for the excitation of auto-oscillations can be achieved.
\Cref{fig:Schematic_AMR}(b) shows a scanning electron micrograph of a typical EP-SHO device.

In this article, we study and compare two types of SHOs: the standard SHO (S-SHO) similar to that shown in \cref{fig:SHO_types}(a) and the EP-SHO. 
In the S-SHO configuration, moderate-amplitude auto-oscillatory dynamics are driven by the antidamping spin Hall torque around the effective magnetic field often dominated by the applied field $\bm{H}$.

The maximum antidamping spin Hall torque efficiency in this configuration is achieved for a saturating  field $\bm{H}$ applied parallel to the direction of the spin Hall current polarization (along the $y$-axis: $\theta = 90^\circ$, $\phiH = 90^\circ$) \cite{hoffmannSpinHall2013}. However, the maximum efficiency of converting magnetization oscillations to resistance oscillations due to AMR oscillations is achieved at $\phi = 45^\circ$. For this reason, the external field is usually applied at an angle between $\phiH = 45^\circ$ and $\phiH = 90^\circ$ as a compromise \cite{chenSpinOrbit2020}. 

In the EP-SHO configuration, the applied field $H$ is nearly zero and the energy landscape is dominated by internal fields: shape anisotropy and PMA. 
The goal is to artificially manufacture an easy-plane in the $xz$-plane, such that spin Hall current from the Pt underlayer is polarized orthogonal to the easy-plane.
In this case, the spin Hall torque pushes the magnetization out of the easy-plane where it precesses about the effective easy-plane field, as shown in \cref{fig:Schematic_AMR}(a).

Magnetic shape anisotropy for a nanowire of rectangular cross section can be approximately described via demagnetization fields along the three principal axes: $\HDx = -4 \pi \Dx \Mx$, $\HDy = -4 \pi \Dy \My$, and $\HDz = -4 \pi \Dz \Mz$, where $D_i$  are the demagnetization factors and $M_i$ are the magnetization components in the $i=x, y, z$-directions. 
The saturation magnetization of the Co$|$Ni superlattice is estimated from thickness dependent FMR measurements to be $\Ms \approx 997$ emu\,cm$^{-3}$ (Supplementary Note S2). 
The demagnetization factors for the Co$|$Ni nanowire used here can are calculated using analytical expressions derived in ref. \cite{aharoniDemagnetizingFactors1998}: $\Dx = 1.4\times 10^{-4}$, $\Dy = 0.121$, and $\Dz=0.879$. 
Upon patterning the nanowire, the $y$-axis becomes a hard magnetic axis with a maximum demagnetization field of $\HDy = 1.52$ kOe, while the $x$-axis has a maximum demagnetization field of only a few Oe. 
The demagnetization field in the direction perpendicular-to-the-film plane is $\HDz = 11.0$ kOe.
The PMA field $\HPMA$ is always in the opposite direction as $\HDz$; therefore to achieve an  easy $xz$-plane, we require the $\HPMA = \HDz = 11.0$ kOe in order to compensate the demagnetization field along the $z$-axis.

To characterize the $\HPMA$ in the magnetic multilayers used here, we made broadband ferromagnetic resonance measurements (FMR)  \cite{montoyaBroadbandFerromagnetic2014} prior to patterning of the multilayers into the nanowire devices. Using these measurements, we adjusted the Co and Ni layer thicknesses such that the sum of Pt$|$Co and Co$|$Ni interfacial PMA contributions at room temperature is less than, but nearly compensates the demagnetization field in the $z$-direction, as described in Supplementary Note S2.
Furthermore, we find the PMA of the multilayer increases by 12\% upon cooling from 295 K to 4.2 K as discussed in the Supplementary Note S3. 
For this reason, the equilibrium direction of magnetization in our nanowire devices at 4.2 K is along the $z$-axis. 
However, this uniaxial anisotropy is small and the dominant anisotropy is the easy-$xz$-plane anisotropy. 
Furthermore, this small $z$-axis uniaxial anisotropy can then be continuously tuned by temperature from easy-$z$-axis to easy-$x$-axis, achieving perfect easy-$xz$-plane anisotropy at the transition temperature. 
In this work, we tune the temperature via Joule heating by the applied direct current.

In this article, we report measurements of a SHO device with an active region length  $ l = 145\,\mathrm{nm}$ made at $T=$ 4.2 K.
\Cref{fig:Schematic_AMR}(c) shows the resistance of the EP-SHO device as a function of in-plane angle $\phiH$ ($\thetaH = 90^\circ$) of applied magnetic field $H = 4\, \mathrm{kOe}$ and a small probe current of $\Idc = 0.1\,\mathrm{mA}$. From measurements of a similar device, we find the magnetoresistance to be due to both AMR and spin Hall magnetoresistance (SMR) \cite{nakayamaSpinHall2013, kimSpinHall2016} with approximately equal contributions.

\bigskip
\noindent \textbf{\myfont Microwave emission experiment.}  

The auto-oscillatory magnetization dynamics in SHO devices are excited by spin Hall torque \cite{andoElectricManipulation2008} from a direct current $\Idc$ exceeding a critical value $I_\mathrm{c}$. These magnetization auto-oscillations give rise to the device resistance oscillations due to AMR and SMR with the amplitude $\dRac$ and to a microwave voltage with the amplitude $V_\mathrm{ac} \sim \Idc \dRac$ \cite{kiselevMicrowaveOscillations2003}. 
 
\textbf{\begin{figure*}[htbp]
\center
 \includegraphics[width= 1.0\textwidth]{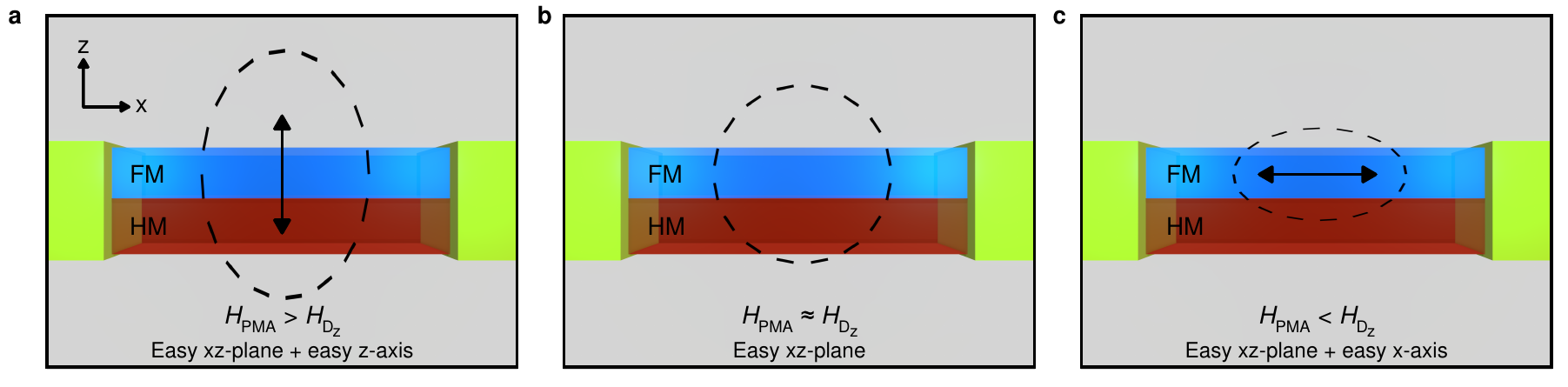}
 \caption{\textbf{Easy plane energy landscape.} Figures show $xz$-plane side view of nanowire device.
 Dashed ellipsoids show constant energy contours of magnetization in the easy $xz$-plane and double-headed arrows indicate easy-axis.
 \textbf{(a)} When the perpendicular anisotropy field $\HPMA$ is larger than the demagnetization field $\HDz$, $z$-axis is an easy axis within the dominant $xz$ easy plane.
 \textbf{(b)} When $\HPMA = \HDz$, perfect easy plane anisotropy is realized in the $xz$ plane. 
 \textbf{(c)} For $\HPMA < \HDz$, $x$-axis is an easy axis within the dominant $xz$ easy plane.
\label{fig:Energy_landscape}}
 \end{figure*}}

We first study the S-SHO configuration shown in \cref{fig:SHO_types}(a) achieved by application of a large magnetic field $H=$ 2.2 kOe in the plane of the sample at $\phiH = 68^\circ$, $\thetaH = 90^\circ$. \Cref{fig:Emission_experiment}(a) shows power spectral density (PSD) measured in this S-SHO configuration as a function of $\Idc$. 
The dynamics show a blue frequency shift with increasing $\Idc$ above the critical current $I_\mathrm{c} =1\,\mathrm{mA}$. 
The observed blue frequency shift is a nonlinear effect expected for the case of a saturating magnetic field applied perpendicular to the easy plane \cite{kiselevCurrentInducedNanomagnet2004, slavinNonlinearAutoOscillator2009}.
\Cref{fig:Emission_experiment}(c) shows the total integrated microwave power $P$ generated by the SHO as a function of $\Idc$. 
The integrated power versus $\Idc$ is non-monotonic and peaks at  $P=74$ pW near $\Idc \approx 2.15 \, \mathrm{mA}$. All values of microwave power given in this article are those delivered a standard $50 \, \Omega$ load.

We next measure the same device in a nearly zero external magnetic field -- a configuration allowing us to achieve the EP-SHO regime of operation.  In the EP-SHO configuration, the energy landscape is dominated by PMA and the shape anisotropy fields.
\Cref{fig:Emission_experiment}(b) shows measured PSD as a function of $\Idc$ for $H = 140\,\mathrm{Oe}$ and $\left( \phiH = 188^\circ \right)$; similar results were found for other small applied field values. 
A low frequency auto-oscillatory mode labelled $m$1 is excited for currents exceeding $\Idc =1.8\,\mathrm{mA}$. 
In contrast to the standard high-field SHO regime, the emission frequency initially red-shifts with increasing $\Idc$ and reaches a minimum of $0.78\,\mathrm{GHz}$ at $\Idc \approx 2.44 \,\mathrm{mA}$.
For $\Idc > 2.44 \,\mathrm{mA}$, the emission frequency blue-shifts.
\Cref{fig:Emission_experiment}(d) shows the integrated power of the SHO in this low-field regime as a function of $\Idc$.
A non-monotonic dependence of microwave emission power is observed with the maximum value of 217 pW reached at $\Idc = 2.39 \, \mathrm{mA}$, near the frequency minimum. 
We also observe 2nd and 3rd harmonics of the mode $m$1, labelled as $m$2 and $m$3 in \cref{fig:Emission_experiment}(b). The presence of the harmonics is indicative of large-amplitude nonlinear oscillations of magnetization.
A higher order mode not harmonically related to $m$1 is observed at frequencies near $m$2 and $m$3.

The non-monotonic SHO frequency dependence on $\Idc$ in the low-field regime of \Cref{fig:Emission_experiment}(b) is due to tuning of $\HPMA$ by Joule heating, which alters the energy landscape in the $xz$-plane as shown in \cref{fig:Energy_landscape}. 
With increasing temperature, the PMA is reduced. 
For $\Idc<2.4 \, \mathrm{mA}$, the perpendicular anisotropy field dominates the $z$-axis demagnetization field $\HPMA > \HDz$ and the energy landscape can be described as a dominant easy $xz$-plane anisotropy with a secondary easy $z$-axis anisotropy within the $xz$-plane, as shown in \cref{fig:Energy_landscape}(a). 
For $\Idc > 2.5\, \mathrm{mA}$, the reduced $\HPMA$ can no longer compensate $\HDz$, and the energy landscape becomes easy-$xz$-plane with a secondary easy $x$-axis, as shown in \cref{fig:Energy_landscape}(c). 
The perfect easy $xz$-plane characterized by $\HPMA = \HDz$ is achieved at the value of $\Idc=2.5\,\mathrm{mA}$ as shown in \cref{fig:Energy_landscape}(b). 
\Cref{fig:Emission_experiment} shows that the microwave power $P = 217$ pW in this EP-SHO regime is significantly enhanced compared to the maximum power $P= 74$ pW in the S-SHO configuration.

The amplitude of resistance oscillations $\dRac$ $(\delta R_{\mathrm{ac}}^{rms} = \dRac / \sqrt{2} )$ shown in Figs. \ref{fig:Emission_experiment} (e) and (f) for the S-SHO and EP-SHO respectively are calculated as \cite{duanNanowireSpin2014}:
\begin{equation}\label{eq:dRac}
    \dRac = \frac{R\left( \Idc \right) + R_{50}}{|\Idc|} \left( \frac{2 P}{R_{50}} \right)^\frac{1}{2},
\end{equation}
where $R_{50} = 50 \, \Omega$ is the load impedance and $R \left( \Idc \right)$ is the resistance of the nanowire at current $\Idc$ (Methods).

\bigskip
\noindent \textbf{\myfont Micromagnetic simulations.} 
Micromagnetic simulations of current-driven magnetization dynamics for both the S-SHO and the EP-SHO configurations are carried out using Mumax3 micromagnetic code \cite{vansteenkisteDesignVerification2014} at $ T = 0\, \mathrm{K}$. Geometry, cell-size, and material parameters used in these simulations are listed in Methods and experimental measurement of material parameters are discussed in Supplementary Note S2. Technical details of the simulations are given in Supplementary Note S3.
\textbf{\begin{figure*}[t]
\center
 \includegraphics[width= 1.0\textwidth]{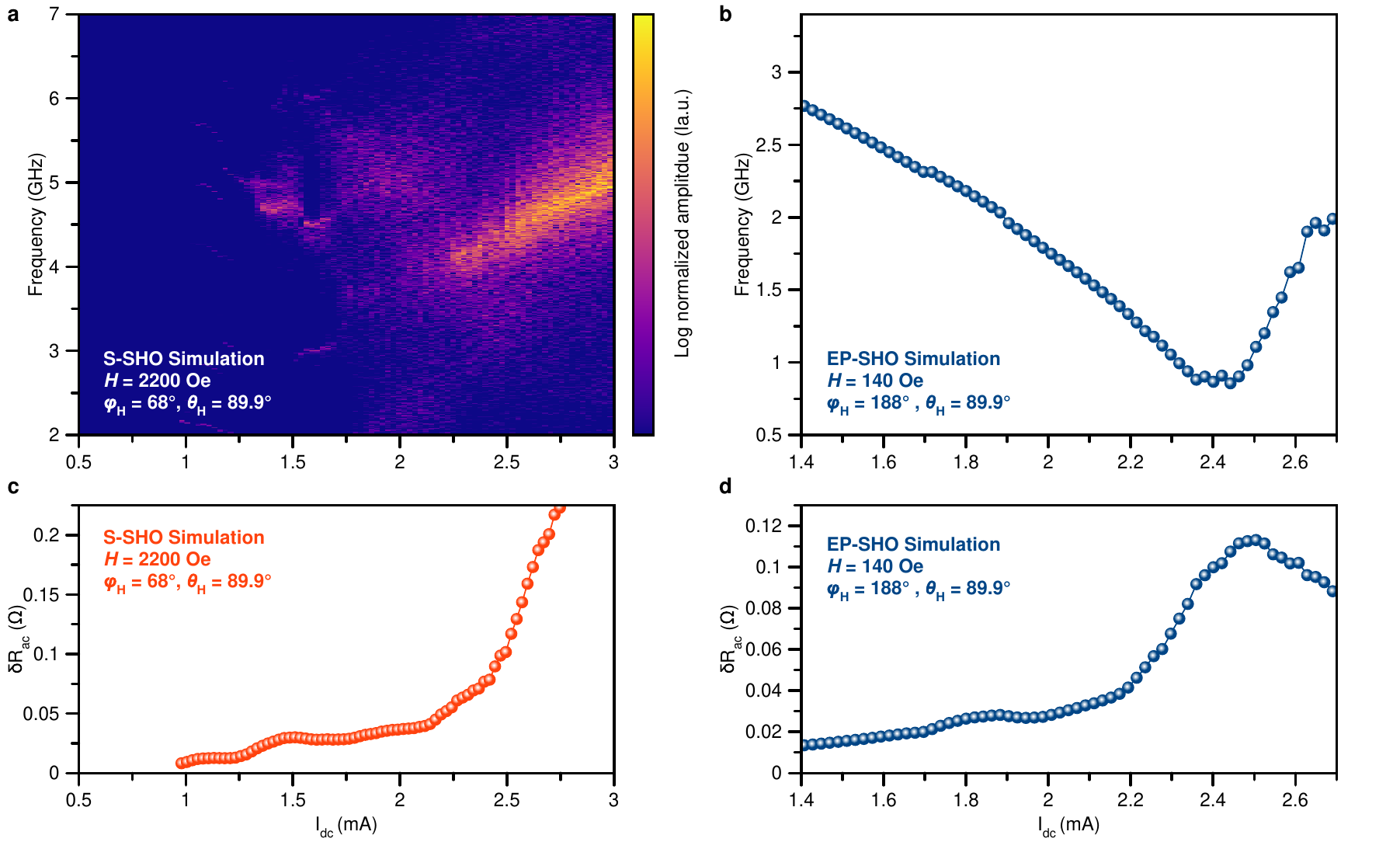}
 \caption{\textbf{Micromagnetic simulation of spin-orbit torque nano-oscillator.}
 Figures show analysis of micromagnetic simulation data of resistance auto oscillations at each current (\textbf{a}) for high field S-SHO configuration (via FFT) (\textbf{b}) for low field EP-SHO configuration (via time domain analysis). 
 Calculated amplitudes of resistance oscillations for (\textbf{c}) high-field S-SHO and (\textbf{d}) low-field EP-SHO.
\label{fig:Emission_sims}}
 \end{figure*}}

Spin Hall torque is applied to the 145 nm long active region in the middle of the nanowire. 
In addition, we account for current-induced Oersted field and Joule heating. 
The Oersted field applied to the FM magnetization in the $-y$ direction arises from electric current in the Pt layer.
In Supplementary Note S3, we show that the magnitude of this field is $66.9 \, \mathrm{Oe}$ per mA $\Idc$.
We also take into account current-induced reduction of $\HPMA$ via Joule heating in the SHO active region. 
Outside the active region, we assume PMA to be equal to its film value at $T=4.2$ K, $\HPMA = 11.7 \,\mathrm{kOe}$. 
This value of PMA results in $z$-axis being the easy axis within the easy-$xz$-plane outside of the active region. 
In the active region, $\HPMA$ is assumed to be a linearly decreasing function of $\Idc$ with the experimentally determined slope of $-494 \, \mathrm{Oe \, mA^{-1}}$ as described in the Supplementary Note S3.
The combination of the Oersted field and reduced $\HPMA$ in the active region creates a magnetic potential well for spin waves, resulting in localization of the auto-oscillatory dynamics to the active region as observed in our simulations.

We first simulate the S-SHO configuration by applying in-plane external field $H = 2.2\,\mathrm{kOe}$ at $\phiH = 68^\circ$ and $\thetaH = 89.9^\circ$. 
The tilt of the external field by $0.1^\circ$ away from the film plane is used to eliminate simulation artifacts possible in highly symmetric systems. 
The system is initialized to uniform magnetization along \phiH \, and then relaxed to its minimum energy state prior to turning on the spin Hall torque. 
We conduct a series of simulations for applied currents in the range from $ \Idc = 0.5 \,\mathrm{mA} $ to $ 3.0 \,\mathrm{mA} $. 
The resulting $x$ and $y$ components of the dynamic magnetization, $m_x \left( t \right)$ and $m_y \left( t \right)$, are used to calculate variation of the sample resistance with time due to AMR and SMR, 
\begin{equation}
    R_\textrm{ac}\left( t \right) =   \Delta R_\textrm{AMR} \langle m_x \left( t \right) \rangle ^{2} - \Delta R_\textrm{SMR} \langle m_y \left( t \right) \rangle ^{2}  ,
\end{equation}
where $ \Delta R_\textrm{AMR} = 0.4 \, \Omega$ is the experimentally measured magnitude of AMR, $ \Delta R_\textrm{SMR} = 0.4 \, \Omega$ is the experimentally measured magnitude of SMR and $\langle ... \rangle$ represents averaging over the active region of SHO.

\Cref{fig:Emission_sims}(a) shows the spectra of the current-driven auto-oscillatory dynamics calculated via fast Fourier transforms (FFT) of $ R_\textrm{ac}\left( t \right) $. 
This figure shows that auto-oscillatory dynamics appears at $\Idc$ exceeding $ 1.25 \,\mathrm{mA} $. 
The magnitude of the resistance oscillations strongly increases when $\Idc$ exceeds $ 2.25 \,\mathrm{mA} $.
For $\Idc>2.25 \,\mathrm{mA} $, the observed auto-oscillatory mode exhibits a nonlinear blue frequency shift.
\Cref{fig:Emission_sims}(c) shows the amplitude of resistance  oscillations $\delta R_\mathrm{ac}$ versus $\Idc$. 

Simulations in the EP-SHO dynamics were made for $ H = 140\, \mathrm{Oe}$,  $\phiH = 188^\circ $, and $\thetaH = 89.9^\circ$.  
These simulations revealed bi-stable behavior of the system in the presence of $\Idc$: at a fixed current above the critical, the system can be either in a dynamic state of large-amplitude magnetization oscillations or in a static state. Example of this behavior is shown in Supplementary Fig. S5.
We thus expect that the system may exhibit random telegraph switching between the dynamic and the static states. While such switching is detrimental to the operation of this device as a coherent microwave source and must be suppressed via design improvements, it may be beneficial for operation of the device as a neuron because small external stimuli result in large-amplitude output voltage spikes \cite{senguptaSpinorbitTorque2015, khymynUltrafastArtificial2018, markovicEasyplaneSpin2022}.
This type of random telegraph noise between large-amplitude dynamics and static states has been previously observed in spin transfer torque oscillators based on nanopillar spin valves \cite{krivorotovTimedomainStudies2008}. 

The bi-stability of the dynamic and static states in the EP-SHO regime warrants the use of time domain analysis described in the Supplementary Note S3 instead of the FFT analysis in order to determine the amplitude of magnetization and resistance oscillations in the dynamic state.
\cref{fig:Emission_sims}(b) shows the bias current dependence of the auto-oscillation frequency determined from this analysis. 
The data reveals a frequency minimum arising from the heating-induced rotation of the easy axis between $z$ and $x$ axes. 
These simulation data are in excellent agreement with the experiment as illustrated by the nearly perfect overlap of the  micromagnetic and experimental data in \cref{fig:Emission_experiment}(b).

\Cref{fig:Emission_sims}(d) shows the amplitude of resistance oscillations $\delta R_\mathrm{ac}$ versus $\Idc$ given by our micromagnetic simulations in this EP-SHO regime (see Supplementary Note S3 for details).
The data show that the amplitude of resistance oscillations is maximized near the frequency minimum where the perfect easy-$xz$-plane anisotropy is realized. This non-monotonic dependence of the amplitude of auto-oscillations on $\Idc$ is expected for the EP-SHO dynamics and is consistent with the experimental data in \cref{fig:Emission_sims}(d).

Supplementary Movie 1 shows spatially resolved time evolution of current-driven magnetization dynamics given by our micromagnetic simulations for $\Idc = 2.44 \, \mathrm{mA}$, which corresponds to the maximum of $\delta R_\mathrm{ac}(\Idc)$.
\Cref{fig:Snapshots}(a)-(c) show three snapshots from this Movie within one period of the auto-oscillations.
\Cref{fig:Snapshots}(a) shows the dynamic micromagnetic state in the active region at $t=20.18$ ns after application of spin Hall torque. 
At this time, the magnetization in the active region points predominantly in the $+z$-direction. 
The magnetization in the active region subsequently precesses towards the $+x$-direction, as shown in \cref{fig:Snapshots}(b) at $t = 20.69 \, \mathrm{ns}$. 
These dynamics are consistent with those expected for an ideal EP-SHO shown in \cref{fig:SHO_types}(b). 
The next expected stage of the ideal EP-SHO dynamics is precession of magnetization towards the $-z$-direction.
Instead of these ideal dynamics, the magnetization of the nanowire EP-SHO rotates towards the $-y$-direction, as shown in \cref{fig:Snapshots}(c) at $t = 20.93$ ns. 
From here, the magnetization precesses towards the $-x$-direction before returning to the $+z$-direction just like in the case of the ideal EP-SHO in \cref{fig:SHO_types}(b). 
This precession cycle repeats with small cycle-to-cycle variations of the micromagnetic states, as may be expected for a nonlinear dynamical system with many degrees of freedom \cite{devolderChaosMagnetic2019}.

Supplementary Movie 2 shows the auto-oscillatory dynamics of the magnetization vector averaged over the EP-SHO active region.
\Cref{fig:Snapshots}(d) displays a frame from this Movie showing the path traced by the averaged magnetization vector for two consecutive cycles of precession. 
\cref{fig:Snapshots}(d), illustrates that large-angle dynamics expected for an EP-SHO are indeed excited by spin Hall torque. However, compared to the ideal EP-SHO, these dynamics are limited to the $+z$ half-space.
This departure from the ideal EP-SHO dynamics is due to exchange coupling to the static magnetization outside of the active region that is magnetized along the $+z$-direction. Supplementary Movie 3 and Supplementary Fig. 6 show the corresponding S-SHO auto-oscillatory dynamics of the magnetization vector averaged over the S-SHO active region.

\textbf{\begin{figure}[tbp]
\center
 \includegraphics[width= 1.0\columnwidth]{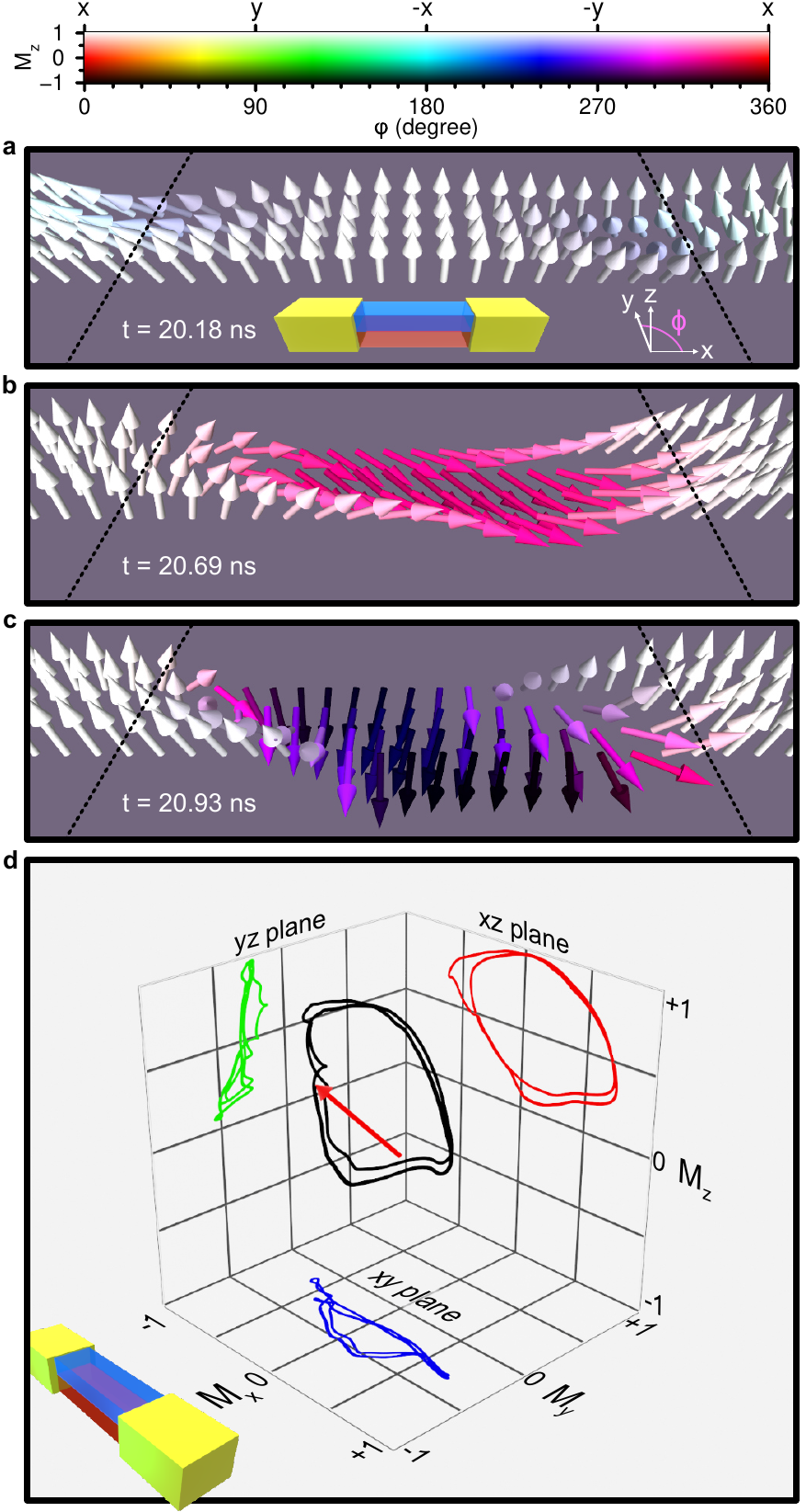}
 \caption{\textbf{Micromagnetic snapshots of EP-SHO auto-oscillations at $\bm{\Idc = 2.44\, \mathrm{mA}} $}. 
 (\textbf{a}) magnetization stars from predominantly $+z$-direction at $t=20.18$\,ns, (\textbf{b}) then precesses towards the $+x$-direction at $t=20.69$\,ns, (\textbf{c}) and subsequently rotates towards the $-y$-direction at $t=20.93$\,ns. 
 Black dashed lines indicate boundaries of the active region. 
 (\textbf{d}) Average magnetization of the active region traced over two periods of auto-oscillations.
\label{fig:Snapshots}}
 \end{figure}}

\bigskip
\begin{large}
\noindent \textbf{\myfont Discussion}
\end{large}

Our experimental data in \cref{fig:Emission_experiment} demonstrate that the easy-plane regime of SHO operation results in a significant power boost compared to the conventional SHO regime. For the nanowire SHO geometry, we observe a power increase by a factor of three in the EP-SHO regime. Micromagnetic simulations of magnetization dynamics for the  EP-SHO and the S-SHO regimes are in qualitative agreement with the experimental observations. 

We find full quantitative agreement between the measured and simulated dependence of the EP-SHO generation frequency as a function of direct current bias $\Idc$ as shown in \cref{fig:Emission_experiment} (b). 
The frequency minimum in these data is observed at the bias current corresponding to the perfect easy-$xz$-plane anisotropy achieved via PMA tuning by Ohmic heating.  

The experimentally measured dependence of the amplitude of resistance oscillations $\dRac$ on $\Idc$ in the EP-SHO regime shown in \cref{fig:Emission_experiment}(f) is qualitatively similar to that given by the simulations in \cref{fig:Emission_sims}(d): in both cases the auto-oscillation amplitude maximum is observed near $\Idc$ corresponding to the perfect easy-$xz$-plane anisotropy. However, the experimentally measured auto-oscillation amplitude is higher than that predicted by the simulations. 
 We attribute this discrepancy to weak exchange coupling between crystallographic grains of the FM film.
 The reason for the auto-oscillation amplitude being limited to the $+z$ half-space in the simulations is strong exchange coupling to static magnetization outside of the active region.
 Therefore, a decrease of exchange coupling to the static magnetization is expected to extend the auto-oscillatory dynamics into the $-z$ half-space, increase $\dRac$ and bring the auto-oscillations closer to the ideal EP-SHO shown in \cref{fig:SHO_types}(b). 
 Recent studies clearly demonstrate significant granularity in Co$|$Ni superlattices deposited by magnetron sputtering and the associated weakening of the inter-grain exchange in such films \cite{capriataImpactRandom2022}.

A recent paper \cite{markovicEasyplaneSpin2022} presented micromagnetic simulations of auto-oscillatory dynamics in a geometry similar to that studied here. 
The auto-oscillatory dynamics found in these simulations are similar to those in an ideal EP-SHO shown in \cref{fig:SHO_types}(b). 
These simulations assume the same value of PMA in the active region and outside of the active region, which is challenging to realize  due to the enhanced Ohmic heating of the active region. 
Our simulations show that inclusion of the enhanced heating-induced PMA reduction in the active region may decrease the amplitude of the auto-oscillations of EP-SHO compared to the ideal case. 

We also find qualitative agreement between the experiment and simulations for the high-field S-SHO regime.
The frequency in the S-SHO regime is found to blue shift with increasing $\Idc$ in the experiment (\cref{fig:Emission_experiment} (a)) and a clear blue frequency shift is seen in the simulations for $\Idc>2.25$\,mA  as shown in \cref{fig:Emission_sims}(a).
This blue shift is a nonlinear dynamical effect expected for a spin torque oscillator with an easy plane magnetic anisotropy and saturating applied magnetic field with a large component perpendicular to the easy plane \cite{kiselevCurrentInducedNanomagnet2004, slavinNonlinearAutoOscillator2009}.

However, the auto-oscillation frequency seen in the experiment is approximately 1 GHz higher than that predicted by the simulations.
This discrepancy is explained by the assumption of ideal magnetic edge of the nanowire used in the simulations: the simulations assume that the FM material properties at the nanowire edge are the same as in the middle of the wire.
This assumption has been previously shown to be incorrect in real devices \cite{mcmichaelEdgeSaturation2006,duanNanowireSpin2014} due to the non-ideal properties of the magnetic edge such as edge roughness \cite{cowburnLateralInterface2000} and magnetic edge dilution \cite{maranvilleVariationThin2007}. 
It has been previously shown that the experimentally measured frequency of spin wave modes in transversely magnetized thin-film nanowires are higher than expected due to the magnetic edge modification \cite{yangParametricResonance2022}.
The magnetic edge modification is also likely to be responsible for deviations of the measured S-SHO frequency from the simulations in the low bias current regime $\Idc<2.25$\,mA. 
The magnetic edge modification has the largest impact on the spin wave frequency for magnetization saturated near the $y$-axis \cite{yangParametricResonance2022}.
This is the reason for a much smaller impact of this effect on auto-oscillation frequency in the EP-SHO regime.

In the high-field regime of S-SHO operation, the simulations predict a continuous increase of the auto-oscillation amplitude $\dRac$ with increasing $\Idc$ up to the largest currents used in the simulations as shown in \cref{fig:Emission_sims}(c). 
In contrast, the experimentally measured $\dRac(\Idc)$ shown in \cref{fig:Emission_experiment} (e) has a maximum near $\Idc=2.15$\,mA. 
The experimentally observed non-monotonic dependence $\dRac(\Idc)$ is consistent with previous studies of S-SHOs \cite{liuMagneticOscillations2012, chenSpinOrbit2020}.
The decrease in the auto-oscillation amplitude in the high current regime has been previously observed in S-SHO nanowire devices and explained \cite{smithDimensionalCrossover2020} via nonlinear magnon scattering \cite{demidovControlMagnetic2011} of the auto-oscillatory mode to thermal magnons. The population of thermal magnons increases in the high-current regime due to the unavoidable Joule heating, resulting in an enhancement of the nonlinear scattering from the auto-oscillatory mode and the associated decrease of its amplitude.

Micromagnetic simulations do not account for thermal magnons, and thus the auto-oscillation amplitude continues to increase with increasing $\Idc$ in the simulations as shown in \cref{fig:Emission_sims}(c). 
It has been demonstrated that nonlinear magnon scattering in S-SHOs increases with increasing ellipticity of the spin wave modes \cite{divinskiyControlledNonlinear2019}. 
Given the nearly easy-$xz$-plane character of anisotropy in our SHO devices, one may expect low ellipticity of spin wave modes and thus low nonlinear scattering rates in the standard mode of the SHO operation when magnetic field is applied along the $y$-axis. 
However, magnetic field in our experiment is applied at a significant angle with respect to the $y$-axis ($\phiH = 68^\circ$) and its magnitude is similar to the easy-$xz$-plane anisotropy field. This results in a significant ellipticity of spin wave modes in the system and turns on the nonlinear scattering to thermal magnons. For this reason, the auto-oscillatory mode amplitude in the high-current regime decreases with increasing $\Idc$ as seen in the experimental data in \cref{fig:Emission_experiment} (e).

Three major pathways to enhance the microwave power output of spin-orbit torque oscillators are: (i) increase the amplitude of magnetization auto-oscillations, (ii) increase conversion efficiency of magnetic oscillations into electric microwave signal and (iii) take advantage of phase locking in arrays of spin torque oscillators to harness phase coherence of their collective dynamics. 
While it is likely that the ultimate future high-power spin-orbit torque oscillator devices will combine all three approaches, an important immediate task is to find optimal solutions to all three individual approaches prior to combining them into a device with the ultimate high-power performance.
It is interesting to note that this problem has been largely solved for spin transfer torque oscillators where the large amplitude of magnetization oscillations is achieved in vortex-based oscillators \cite{pribiagMagneticVortex2007}, high conversion efficiency is achieved via tunneling magneto-resistance (TMR) in MTJs \cite{tsunegiMicrowaveEmission2016} and phase locking of several of vortex oscillators has been demonstrated \cite{ruotoloPhaselockingMagnetic2009, tsunegiScalingElectrically2018}. 
Achieving this degree of success is a grand challenge for spin-orbit torque oscillators. If realized, this goal can lead to high-power spin-orbit torque oscillator devices that are more energy-efficient than spin transfer torque oscillators and operate at higher microwave frequencies than vortex-based oscillators.

Our experimental demonstration of an EP-SHO solves the problem of achieving large-amplitude auto-oscillations in a single spin-orbit torque oscillator. 
A common approach to increasing the amplitude of magnetic auto-oscillations in spin transfer torque devices is excitation of auto-oscillations of a magnetic vortex \cite{pribiagMagneticVortex2007}.
However, vortex oscillators driven by spin Hall torque have not been realized due to the direction of the current polarization being in the FM$|$HM bilayer plane. 
The artificial easy-plane approach shown to work in this paper presents a practical solution for large-amplitude SHO devices.

Recently, tunable PMA in a SHO based on a Pt$|$Co$|$Ni multilayer was used to decrease the detrimental nonlinear magnetic damping via minimizing the ellipticity of magnetization precession \cite{divinskiyControlledNonlinear2019}. 
This SHO based on a 0.5 $\mu$m diameter disc was shown to operate with small nonlinear damping in the standard high-field SHO regime.
However, the disc geometry does not support the artificial easy-$xz$-plane anisotropy demonstrated in this work.

High conversion efficiency of magnetic auto-oscillations into electric microwave signal can be achieved in SHO devices with high magnetoresistance. 
To this end, the most promising approach is SHOs utilizing TMR, such as 3-terminal devices where a nanoscale MTJ is patterned on top of the HM material \cite{liuMagneticOscillations2012, jueComparisonSpintransfer2018}. 
In such SHOs, the drive and the readout currents can be separately controlled, which allows for low power consumption combined with high output power.
Another promising approach to boosting SHO output power while keeping Ohmic losses low utilizes current-in-plane giant magnetoresistance in a 2-terminal device \cite{chenSpinOrbit2020}.
This approach takes advantage of the identical angular symmetries of spin Hall torque giant magnetoresistance to simultaneously maximize the amplitude of resistance oscillations and spin Hall torque efficiency.

Finally, phase locking in one- and two-dimensional arrays of SHOs \cite{zahedinejadMemristiveControl2022,zahedinejadTwodimensionalMutually2020} has been experimentally demonstrated to significantly boost the SHO output power. 
Therefore, with the addition of the present work, all three individual components needed for making high-power SHO oscillator systems have been experimentally demonstrated. 
We thus expect that integrated SHO devices capable of generating the ultimate high microwave power are now within reach.

In conclusion, our work provides the first experimental realization of an easy-plane spin Hall oscillator. This oscillator can operate without a bias magnetic field and generate high output microwave power due to the large-amplitude of resistance auto-oscillations excited by spin Hall torque. 
The easy plane magnetic anisotropy perpendicular to the film plane is engineered via tuning the nanowire shape anisotropy and interfacial perpendicular magnetic anisotropy. Our micromagnetic simulations of the oscillator performance are in good qualitative agreement with the measurements. 
Our results set the stage for the development of artificial spiking neuron driven by spin Hall torque \cite{markovicEasyplaneSpin2022} and for further enhancement of the oscillator output power via integration with a tunneling magnetoresistance readout \cite{liuMagneticOscillations2012}.

\bigskip
\begin{large}
\noindent \textbf{\myfont Methods}
\end{large}

\noindent \textbf{\myfont Sample description.} 
The multilayer films were deposited by dc magnetron sputtering on Al$_2$O$_3$(0001) substrates in 2 mTorr of Ar process gas. 
Highly resistive, amorphous Ta seed layer was used to reduce film roughness and absorb spin Hall current from Pt propagating opposite to the Co$|$Ni superlattice. 
The highly resistive Ta cap was used to prevent oxidation of the Co$|$Ni. 
The multilayers were patterned into 50 nm wide, 40 $\mu$m long nanowires by means of electron-beam lithography using DOW-Corning HSQ negative resist and Ar ion mill etching. 
The electrical leads to the nanowire were patterned via electron-beam lithography using a methyl methacrylate/poly(methyl methacrylate) positive resist bilayer followed by the sputter deposition of Ta(5 nm)/Au(40 nm)/ Ta(5 nm) and liftoff. 
The spacing between the leads defined a SHO active region ranging in length from 50\,nm to 450\,nm long in the central part of the nanowire.

\noindent \textbf{\myfont Microwave emission measurements.} 
The microwave power emitted from the SHO was detected using a standard circuit based on microwave spectrum analyser \cite{kiselevMicrowaveOscillations2003}. 
A direct current $\Idc$ was applied to the sample through the low-frequency port of a bias tee. 
The signal from the SHO was amplified by a low-noise microwave amplifier with 45\,dB gain, applied to the high-frequency port of the bias tee and recorded by the microwave spectrum analyzer. 
For these measurements, the sample was placed in a He flow cryostat at a bath temperature of $T=4.2$\,K. 
The values of the microwave power reported here are those delivered to a 50$\,\Omega$ load with the frequency-dependent circuit attenuation and amplification calibrated out. Resistance oscillations are calculated by treating the SHO as a mismatched microwave generator connected to a $50 \, \Omega$ transmission line terminated with matched load (spectrum analyzer) \cite{pozarGeneratorLoad2012}.

\noindent \textbf{\myfont Micromagnetic simulations.}
Micromagnetic simulations were made using the Mumax3 software.
We simulate a 4 $\mu$m $\times$ 50 nm $\times$ 5.85 nm  ferromagnetic nanowire composed of $2048 \times 16 \times 1 $ micromagnetic cells representing the length, width, and thickness, respectively.  
The simulations were made using the experimentally determined material parameters of the Co$|$Ni superlattice: saturation magnetization $ M_\mathrm{s} = 997$ emu\,cm$^{-3}$, Gilbert damping $\alpha=0.027$, and Land\'{e} g-factor $g = 2.18$.
Exchange constant $ A_\mathrm{ex} = 1 \times 10^{-6} \, \mathrm{erg \, cm^{-1}}$ and spin Hall angle $\theta_\mathrm{SH} = 0.07$ were used. 
Constant PMA of $\HPMA = 11.7 \, \mathrm{kOe}$ was used outside of the SHO active region. Current-dependent PMA was used in the SHO active region to capture the effects of Joule heating on anisotropy (see Supplementary Note S3 for details).

\bigskip
\begin{large}
\noindent \textbf{\myfont Data availability}
\end{large}

\noindent All data generated or analysed during this study are included in this published article and are available from the corresponding author on reasonable request.

\bigskip
\begin{large}
\noindent \textbf{\myfont Acknowledgements}
\end{large}

\noindent This work was supported by the National Science Foundation through Awards No. EFMA-1641989 and No. ECCS-2213690. We also acknowledge support by the Army Research Office through Award No. W911NF-16-1-0472. The authors acknowledge the use of facilities and instrumentation at the UC Irvine Materials Research Institute (IMRI), which is supported in part by the National Science Foundation through the UC Irvine Materials Research Science and Engineering Center (DMR-2011967). The authors also acknowledge the use of facilities and instrumentation at the Integrated Nanosystems Research Facility (INRF) in the Samueli School of Engineering at the University of California Irvine. We thank NVIDIA Corporation for the donation of the Tesla K40C GPU used for some of the calculations. This work utilized the infrastructure for high-performance and high-throughput computing, research data storage and analysis, and scientific software tool integration built, operated, and updated by the Research Cyberinfrastructure Center (RCIC) at the University of California, Irvine (UCI). The RCIC provides cluster-based systems, application software, and scalable storage to directly support the UCI research community.

\bigskip
\begin{large}
\noindent \textbf{\myfont References}
\end{large}

\bibliography{refer}

\bigskip
\begin{large}
\noindent \textbf{\myfont Author contributions}
\end{large}

\noindent 
E.A.M developed and deposited the film structures, performed film level measurements, and analyzed film level data. 
E.A.M, C.S., and A.S. developed the nanofabrication process and made the nano-devices. 
C.S. and E.A.M. performed device level measurements and analyzed device data. 
A.K. performed micromagnetic simulations. 
I.N.K. formulated and supervised the study. 
E.A.M., I.N.K. and A.K. wrote the manuscript. All authors discussed the results.

\bigskip
\begin{large}
\noindent \textbf{\myfont Competing interests}
\end{large}

\noindent The authors declare no competing interests.

\end{document}

\title{\myfont Supplementary Material for Easy-plane spin Hall oscillator}

\author{\myfont Eric Arturo Montoya}
\email{eric.montoya@uci.edu}
\affiliation{\myfont Department of Physics and Astronomy, University of California, Irvine, California 92697, USA}
\affiliation{\myfont Department of Physics and Astronomy, University of Utah, Salt Lake City, Utah 84112, USA}

\author{\myfont Amanatullah Khan}
\affiliation{\myfont Department of Physics and Astronomy, University of California, Irvine, California 92697, USA}

\author{\myfont Christopher Safranski}
\affiliation{\myfont Department of Physics and Astronomy, University of California, Irvine, California 92697, USA}

\author{\myfont Andrew Smith}
\affiliation{\myfont Department of Physics and Astronomy, University of California, Irvine, California 92697, USA}

\author{\myfont Ilya N. Krivorotov}
\email{ ilya.krivorotov@uci.edu}
\affiliation{\myfont Department of Physics and Astronomy, University of California, Irvine, California 92697, USA}

\date{\today}

\keywords{}
\maketitle

\makeatletter
\def\@hangfrom@section#1#2#3{\@hangfrom{#1#2}#3}%\MakeTextUppercase{#3}}
\def\@hangfroms@section#1#2{#1#2}%\MakeTextUppercase{#2}}
\makeatother

\section*{Supplementary Note 1: Multilayer film deposition}\label{sec:films}

\begin{figure}[!htb]
\center
 \includegraphics[width= 0.95\linewidth]{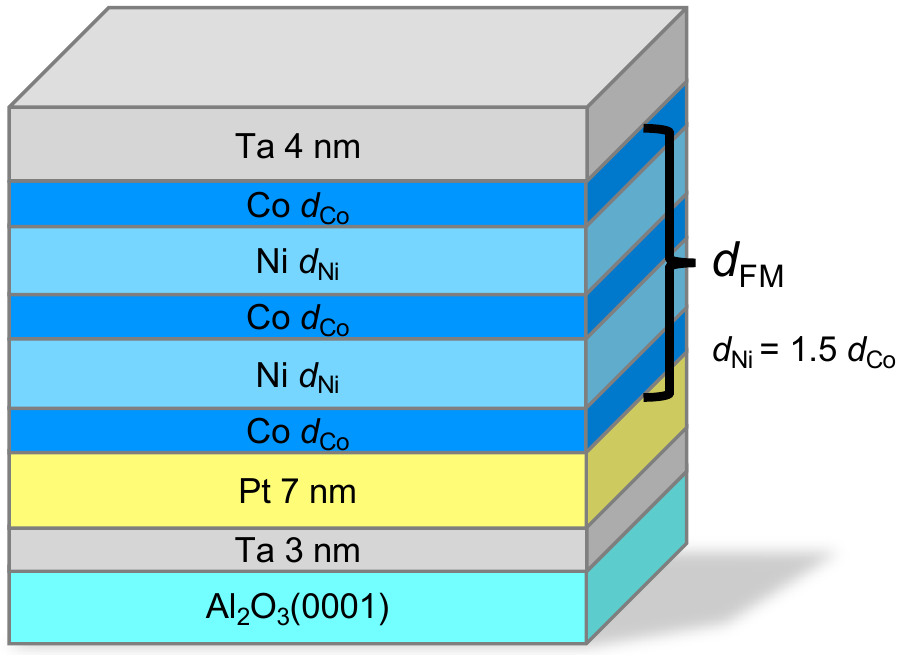}
 \caption{\textbf{Multilayer film used in this work.} FM layer is a Co$|$Ni supperlattice $\left[\mathrm{Co}(\dCo)|\mathrm{Ni}(\dNi) \right]_{2}|\mathrm{Co}(\dCo)$, where $\dNi=  1.5 \, \dCo$ are the thickness of the individual Ni and Co layers.
 The FM thickness $\dFM$ is given by the total thickness of the Co$|$Ni supperlattice.  Note: individual layer thicknesses are not to scale.
\label{fig:film}}
 \end{figure}

Multilayer films were deposited by magnatron sputtering on Al$_2$O$_3 (0001)$ substrates at room temperature in 2 mTorr Ar.
The base pressure of the deposition system was below $3.0 \times 10^{-8} \, \mathrm{Torr}$. Supplementary Fig. \ref{fig:film} shows a cartoon representation of the multilayer films used in this paper. 
A 3 nm Ta seed layer was used for all samples to promote smooth multilayer growth \cite{montoyaSpinTransport2016} and 4 nm Ta capping layer was used to prevent oxidation of the ferromagnetic layers. 
A 7 nm thick Pt layer served as the spin Hall heavy metal (HM) for all samples. 

The ferromagnetic metal (FM) was a Co$|$Ni superlattice $\left[\mathrm{Co}(\dCo)|\mathrm{Ni}(\dNi) \right]_{2}|\mathrm{Co}(\dCo)$, where $\dCo$ and $\dNi$ are the thickness of the individual Co and Ni layers. 
We fix the individual relative Co and Ni thicknesses to $\dNi = 1.5 \, \dCo $ and vary the total superlattice FM thickness $\dFM$ to determine magnetic anisotropy and damping properties at the film-level as described below. 
Note the total thickness ratio of Co:Ni is 1:1, as there are 3 Co and 2 Ni layers in total.
Using the anisotropy and damping optimization described in the next section, we select the final film stack of Al$_2$O$_3 (0001)||$Ta (3 nm)$|$Pt (7 nm)$|\left[\mathrm{Co}(0.98\,\mathrm{nm})|\mathrm{Ni}(1.46\,\mathrm{nm}) \right]_{2}|\mathrm{Co}(0.98\,\mathrm{nm})|$Ta (4 nm) to pattern into nanowires as described in the Methods section of the main article. The total FM thickness is $\dFM = 5.85$ nm.

\section*{Supplementary Note 2: Broadband ferromagnetic resonance}\label{FMR}

We employed a conventional broadband ferromagnetic resonance (FMR) technique to measure magnetic anisotropy and magnetic damping at the film level \cite{kalarickalFerromagneticResonance2006}. All film-level FMR measurements were performed at discrete frequencies in a field swept, field modulated configuration using a broadband microwave generator, coplanar waveguide, planar-doped detector diode, and lock-in amplifier, as detailed in Ref.\cite{montoyaBroadbandFerromagnetic2014}. 
The measured FMR data are described by an admixture of the $\chi^{'}$ and $\chi^{''}$ components of the  complex transverse magnetic susceptibility, $ \chi = \chi^{'} + i \chi^{''} $. 
The FMR data are fit as described by Ref.~\cite{montoyaBroadbandFerromagnetic2014}.

The FMR magnetization dynamics are well described by the Landau-Lifshitz-Gilbert (LLG) equation:
\begin{equation}\label{eq:LLG}
 \frac{ \partial \bm{M} }{ \partial t } = -\gamma \left[ \bm{M} \times \bm{H}_{\mathrm{eff}}  \right]  +  \alpha \left[ \bm{M} \times \frac{ \partial \uvec{m} }{ \partial t }  \right],
\end{equation}	
where $ \bm{M} $ is the instantaneous magnetization vector with magnitude  $ M_{\rm{s}} $, $ \uvec{m} $ is the unit vector parallel to $ \bm{M} $, $\bm{H}_{\rm{eff}}$ is the sum of internal and external magnetic fields, $\gamma = g \mu_{\mathrm{B}}/\hbar $ is the absolute value of the gyromagnetic ratio, $g$ is the Land\'{e} g-factor, $\mu_{\mathrm{B}}$ is the Bohr magneton, $\hbar$ is the reduced Planck constant, and $\alpha$ is the dimensionless Gilbert damping parameter.

\subsection{A. Interfacial perpendicular magnetic anisotropy}\label{PMA}

The Co$|$Ni supperlattice films are polycrystalline and were found to display negligible in-plane anisotropy; therefore, the in-plane ferromagnetic resonance condition is well described by:
\begin{equation}\label{eq:Resonance}
\left( \frac{\omega}{\gamma} \right)^2 = \left( H_{\mathrm{FMR}} \right)\left( H_{\mathrm{FMR}} + 4 \pi M_{\mathrm{eff}} \right),
\end{equation}	
where $\omega =2 \pi f $ is the microwave angular frequency, $H_{\mathrm{FMR}}$ is the resonance field, 
\begin{equation}\label{eq:SatEff}
     4\pi M_{\mathrm{eff}} = 4 \pi \Dz \Ms - \HPMA,
\end{equation}
where $\Ms$ is the saturation magnetization, $\Dz$ is the demagnetization factor for the $z-$axis, and $\HPMA$ is the uniaxial perpendicular anisotropy field \cite{kardaszSpinDynamics2011}. For an extended ultrathin film, $\Dz \simeq 1$ \cite{heinrichUltrathinMetallic1993}. The first term on the right-hand-side of Supplementary Fig. \ref{eq:SatEff} represents the demagnetization field that is present when the extended film magnetization direction lies perpendicular-to-film-plane ($z-$axis). The demagnetization field acts to force the magnetization in the film plane and is proportional and opposite in direction to the $z-$component of magnetization, $\HDz = 4 \pi \Dz \Mz$. $\HPMA$ acts opposite to the $z-$axis demagnetization field, forcing the magnetization perpendicular-to-film-plane. We find that in our samples the bulk contribution to PMA is negligible and that the interfacial contributions from the Co$|$Pt and Co$|$Ni interfaces dominate \cite{aroraOriginPerpendicular2017}. The total uniaxial perpendicular anisotropy field is given by \cite{kardaszSpinDynamics2011}:
\begin{equation}\label{eq:HPMA}
    \HPMA = \frac{2 \Kus}{\Ms \dFM},
\end{equation}
where $ \Kus $ is the total interfacial perpendicular-to-film-plane uniaxial anisotropy energy. 

\begin{figure}[!tbp]
\center
 \includegraphics[width= 0.95\linewidth]{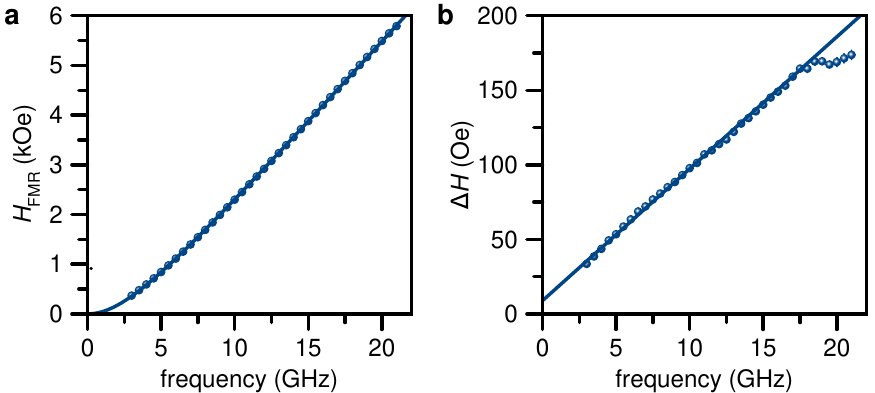}
 \caption{\textbf{Ferromagnetic resonance field and linewidth.}  Ferromagnetic resonance (\textbf{a}) field $\Hfmr$ and (\textbf{b}) linewidth $\DHfmr$ as a function of frequency.
\label{fig:Hfmr_DH}}
 \end{figure}

\begin{figure}[!tbp]
\center
 \includegraphics[width= 0.95\linewidth]{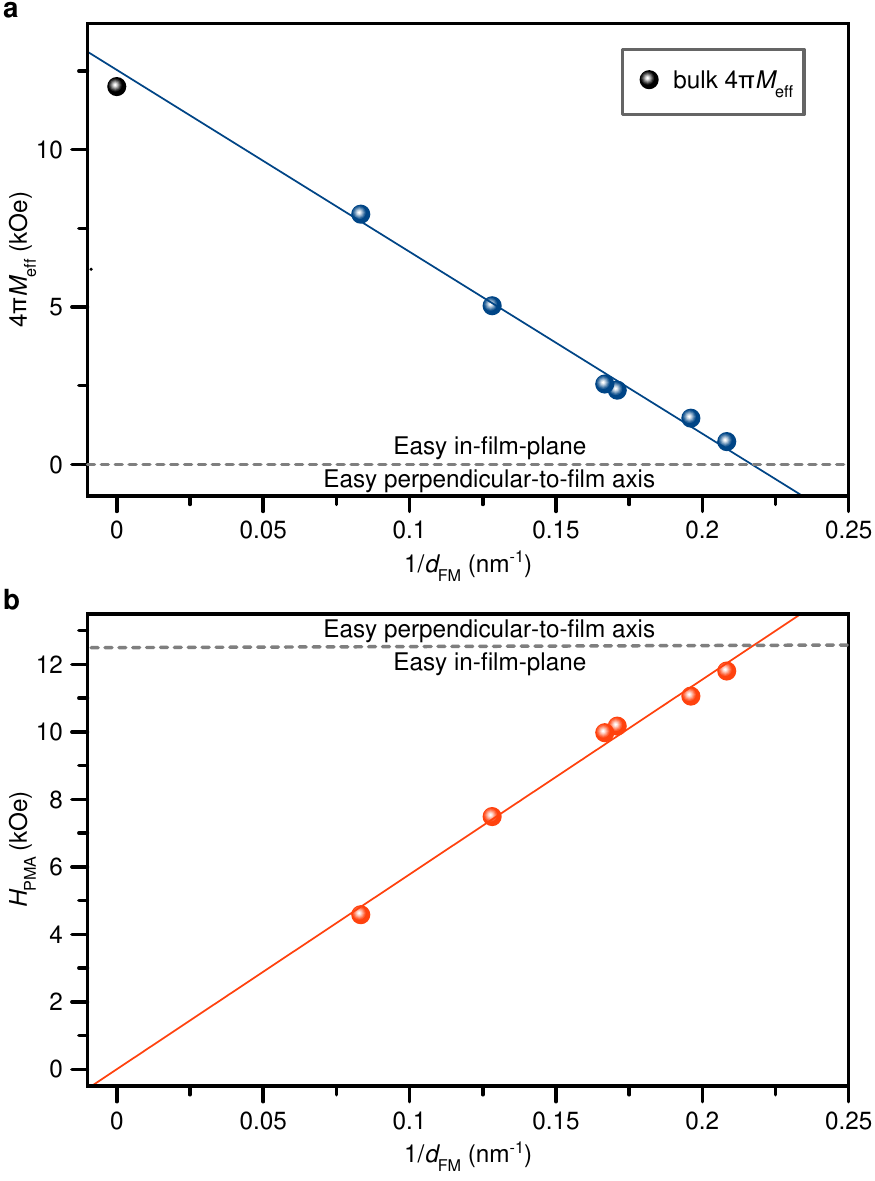}
 \caption{\textbf{Magnetic anisotropy versus inverse ferromagnetic film thickness.} (\textbf{a}) Effective saturation induction $4 \pi \Meff$ and (\textbf{b}) corresponding perpendicular anisotropy field $\HPMA$  as a function of inverse FM thickness $1/\dFM$. A positive $4 \pi \Meff$ indicates a magnetic easy plane coinciding with the film plane, while a negative value indicates a magnetic easy perpendicular-to-film axis.
\label{fig:HPMA_vs_deff}}
 \end{figure}

First we used room-temperature FMR measurements of the dependence of $4 \pi M_{\mathrm{eff}}$ on FM thickness $\dFM$ to quantify the perpendicular anisotropy.
Supplementary Fig.~\ref{fig:Hfmr_DH}(a) shows example data of $H_{\mathrm{FMR}}$ as a function of frequency for the film used to make the device described in the main article. 
The data are fit using equation~\ref{eq:Resonance} to extract  $4 \pi M_{\mathrm{eff}}$ and $g$; the solid line in Supplementary Fig.~\ref{fig:Hfmr_DH}(a) is the resulting fit. 
We find that all samples have $g \simeq 2.18$.
Supplementary Fig.~\ref{fig:HPMA_vs_deff}(a) shows $4 \pi M_{\mathrm{eff}}$ as a function of the inverse FM film thickness $1/\dFM$. 
The solid line is a fit using Supplementary Fig. \ref{eq:SatEff} and \ref{eq:HPMA} with $\Dz = 1$, which is used to extract the magnetic material parameters $4 \pi \Ms = 12.5 \, \mathrm{kOe} $ $(\Ms = 997 \mathrm{\, emu \, cm}^{-3})$ and $\Kus = 2.9 \, \mathrm{erg} \, \mathrm{cm}^{-2}$. 
We find that experimentally determined saturation induction is in good agreement with an expected value of a Co$|$Ni superlattice $4 \pi \Ms = 12.0 \, \mathrm{kOe}$ $(\Ms = 953\, \mathrm{Oe})$, which is estimated by taking volume weighted average of bulk $\Ms$ for Co and Ni: 1422 Oe  and 484 Oe, respectively \cite{cullityIntroductionMagnetic2009}. Note the thickness ratio 1:1 for Co:Ni gives a volume ratio of 1:1; therefore, the estimated $\Ms$ is simply the average of that of Co and Ni in our particular case.
We have plotted the bulk-like value as the horizontal dashed line in Supplementary Fig. \ref{fig:HPMA_vs_deff}(a). Supplementary Fig. \ref{fig:HPMA_vs_deff}(b) shows $\HPMA$ as a function of $1/\dFM$.

\subsection{B. Magnetic damping}\label{damping}

The measured FMR linewidth defined as half-width of the resonance curve \cite{montoyaBroadbandFerromagnetic2014} is well described by Gilbert-like damping,
\begin{equation}\label{eq:GilbertDamping}
\Delta H(\omega)= \alpha \frac{\omega}{\gamma} + \Delta H(0),
\end{equation}	
where $ \Delta H(0) $ is the zero-frequency line broadening due to long-range magnetic inhomogeneity \cite{heinrichUltrathinMetallic1993, mcmichaelLocalizedFerromagnetic2003}.

 Supplementary Fig. \ref{fig:Hfmr_DH}(b) shows data of $\Delta H$ as a function of frequency for the $\dFM =5.85$ nm film used to make the devices in the main article. 
 The data are fit using equation~\ref{eq:GilbertDamping} to extract the parameters $\alpha = 2.7 \times 10^{-3}$ and $ \Delta H(0) = 9 \, \mathrm{Oe}$.
 This damping parameter is used in the micromagnetic simulations.

\subsection{C. Temperature dependence of perpendicular anisotropy}\label{sec:PMA_vs_T}

\begin{figure}[!tbp]
\center
 \includegraphics[width= 0.95\linewidth]{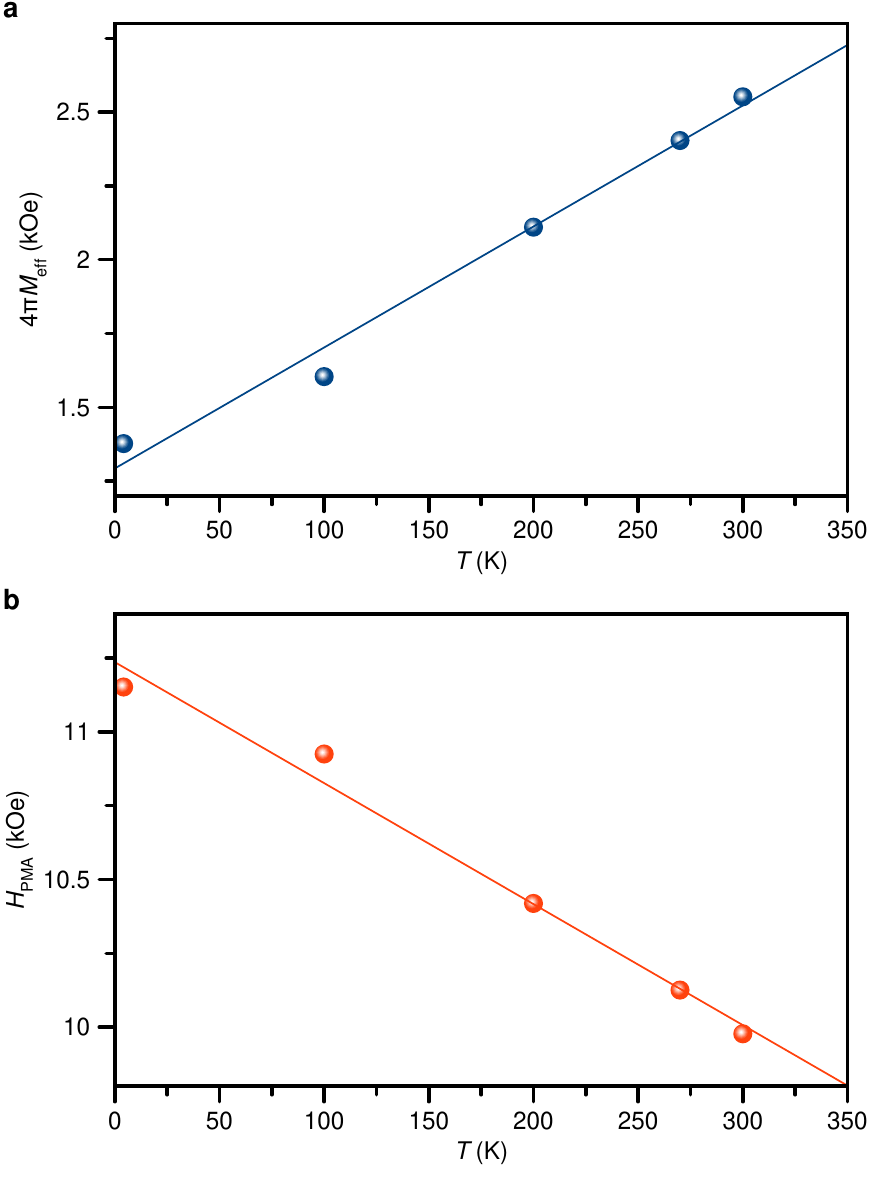}
 \caption{\textbf{Temperature dependence of magnetic anisotropy.} (\textbf{a}) Effective saturation induction $4 \pi \Meff$ and (\textbf{b}) corresponding perpendicular anisotropy field $\HPMA$  as a function of bath temperature $T$.
\label{fig:HPMA_vs_T}}
 \end{figure}

We carried out FMR measurements in a continuous flow $^4$He cryostat to determine the temperature dependence of perpendicular anisotropy in the Co$|$Ni superlattice.
Supplementary Fig. \ref{fig:HPMA_vs_T}(a) shows the saturation induction $4 \pi \Meff$ for a $\dFM = 6 \, \mathrm{nm}$ film as a function of bath temperature $T$. Supplementary Fig. \ref{fig:HPMA_vs_T}(b) shows the corresponding change in uniaxial perpendicular anisotropy field $\HPMA (T)$. 
We find that the perpendicular anisotropy increases approximately linearly in temperature by $12 \%$ upon cooling from $295\, \mathrm{K}$ to $4.2\, \mathrm{K}$. 
We additionally measured device resistance $R$ as a function of bath temperature $T$ at a small probe bias current of $\Idc = 0.1$ mA.
We subsequently use these $R(T)$ data to estimate the nanowire device temperature from its resistance ($T(R)$) when large $\Idc$ is applied to the device resulting in significant Ohmic heating ($R(\Idc)$) (in e.g. SHO microwave emission measurements).
Combining the measurements of $\HPMA (T)$, $T(R)$ and $R(\Idc)$, we establish the bias current dependence of the perpendicular anisotropy field $\HPMA (\Idc)$ used in the micromagnetic simulations of current-induced magnetization auto-oscillations in nanowire SHOs.

\section*{Supplementary Note 3: Micromagnetic simulations}\label{MuMax}

\subsection{A. Device geometry and materials parameters}
Micromagnetic simulations are carried out using MuMax3 micromagnetic simulator \cite{vansteenkisteDesignVerification2014} at zero temperature ($ T = 0\, \mathrm{K}$) with experimentally determined material parameters: saturation magnetization $ M_\mathrm{s} = 997$ emu\,cm$^{-3}$, Gilbert damping $\alpha=0.027$, and Land\'{e} g-factor $g = 2.18$. We assume an exchange constant $ A_\mathrm{ex} = 1 \times 10^{-6} \, \mathrm{erg/cm}$ \cite{beaujourFerromagneticResonance2007}.

The simulation geometry consists of a nanowire with total length $l = 4 \, \mu$m, width $w = 50 \, \mathrm{nm}$, and thickness  $\dFM = 5.85 \, \mathrm{nm}$. Spin Hall torque is applied to the 145 nm long active region at the center of the nanowire lenghtwise. The micromagnetic system is composed of $2048 \times 16 \times 1 $ cells, yielding cell sizes of $1.95 \times 3.13 \times 5.85 \, \mathrm{nm}^3$. At the lengthwise ends of the wire we have implemented absorbing boundary conditions via ramping up the damping parameter. We conducted preliminary relaxation simulations with varying $\Kus$ to determine the value in which the nanowire easy axis transitions to out of plane. We found that the relaxed magnetization of the entire nanowire undergoes transition from in plane to out of plane from $\Kus = 2.895 \, \mathrm{erg \, cm^{-2}} \,(\HPMA = 9.927 \, \mathrm{kOe})$ to $\Kus =3.050 \, \mathrm{erg \, cm^{-2}} \,(\HPMA = 10.46 \, \mathrm{kOe})$, which agreed well with our experimental results. The perpendicular magnetic anisotropy outside the active region and in the active region at zero bias current is taken to be $\Kus = 3.4 \, \mathrm{erg \, cm^{-2}} \,(\HPMA = 11.7 \, \mathrm{kOe})$, resulting in an easy-$z$-axis similar to the experiment.

\subsection{B. Current induced effects}
The spin Hall torque resulting from the charge current flowing in the Pt layer is calculated using the Slonczewski spin torque solver available within the Mumax3 software.
The spin Hall torque magnitude is set by the value of spin Hall angle $\theta_\mathrm{SH} = 0.07$ and electric current density flowing in the Pt layer. The spin Hall angle was tuned to the value $\theta_\mathrm{SH} = 0.07$ to match the experimental onset of auto-oscillations for low and high field SHO operation. This value matches well the intrinsic value measured by \citet{wangDeterminationIntrinsic2014} ($\theta_\mathrm{SH} = 0.068$) and is within the range measured for sputtered Pt thin films \cite{sagastaTuningSpin2016}. For positive $\Idc$, the spin current is polarized in the $-y$ direction. The charge current density in Pt is calculated from the wire cross sectional area and the estimated fractional current flow through Pt, which for similar nanowires was found to be $I_\mathrm{dc}^\mathrm{Pt}=0.7 \Idc$  \cite{safranskiSpinOrbit2019}.

We also include an Oersted field $H_\mathrm{Oe}(z)$ parallel to the $y$ axis that scales with charge current in Pt $I_\mathrm{dc}^\mathrm{Pt}$.  
We estimate the Oersted field applied to the FM layer as that produced by a thin Pt ribbon with uniform sheet current density, taking the $y$-axis component \cite{gauthierMagneticField1998}:
 \begin{multline}
    H_\mathrm{Oe}\left( y,z \right) = - \frac{0.7 I_\mathrm{dc}}{w}\frac{ z }{2 \pi  |z|}  \biggl( \text{arctan}\left(\frac{\frac{w}{2}-y}{|z|}\right) \\ + \text{arctan}\biggl( \frac{\frac{w}{2}+y}{|z|}\biggr) \biggr) \left( \frac{4 \pi}{1000} \frac{\mathrm{Oe \, m}}{A} \right) .
    \end{multline}
Here the origin of the $y$- and $z$-axes are taken to be the geometric center of the Pt layer cross-section. The coordinate system matches that in the main text (see main text Fig. 2(a)): $x$-axis along the length of the wire (positive $\Idc$ flows in positive $x$-direction), $z$-axis normal to the substrate surface, and $y$-axis in-plane and perpendicular to the nanowire.

In the simulations, we apply a uniform $H_\mathrm{Oe}$ in the active region taken from $H_\mathrm{Oe}\left( y,z \right)$ averaged over the cross-section of the FM wire:
\begin{equation}
    H_\mathrm{Oe}=\frac{1}{w \dFM}\int _{\frac{\dPt}{2}}^{\dFM+\frac{\dPt}{2}}\int _{-\frac{w}{2}}^{\frac{w}{2}} H_\mathrm{Oe}\left( y,z \right) dy dz .
\end{equation}
We therefore set the Oersted field in the active region to scale with $\Idc$ as  $66.9 \, \mathrm{Oe}$ per mA. 
From the temperature dependent anisotropy and resistivity data we determined that $\HPMA$ $(\Kus)$   shifts by $-493 \, \mathrm{Oe}$ $(-0.1438 \, \mathrm{erg \, cm^{-2}})$  per mA for $\Idc$ near the magnetic reorientation transition. 
The combination of the Oersted field and current-dependent  $\HPMA$ $(\Kus)$ in the active region creates a magnetic potential well for spin waves, resulting in the localization of auto-oscillations within the active region.

\subsection{C. EP-SHO dynamic to static state transition.}
\begin{figure}[!tbp]
\center
 \includegraphics[width= \linewidth]{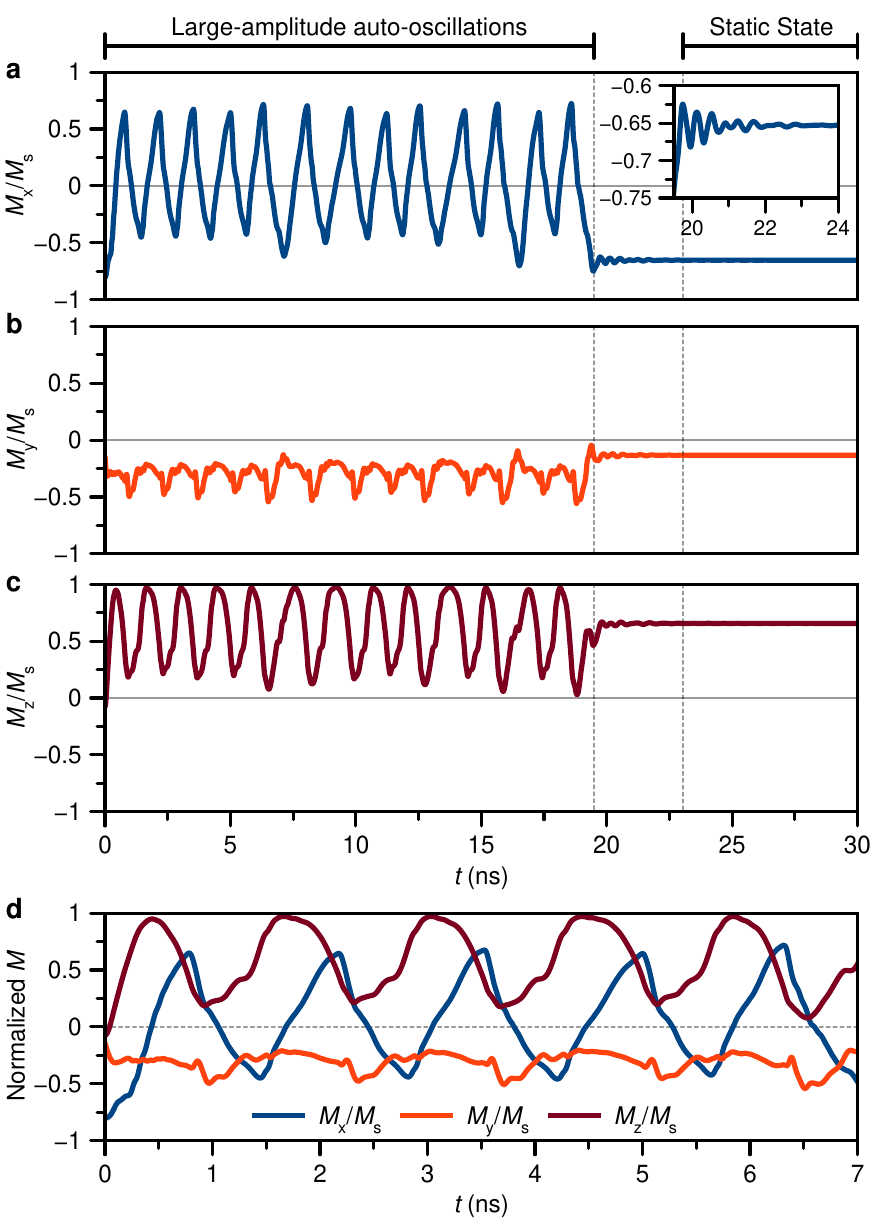}
 \caption{\textbf{Dynamic to static state transition in simulations of the EP-SHO regime.} time dependence of the normalized, averaged over the active region components of the FM layer magnetization (\textbf{a}) $m_x$, (\textbf{b}) $m_y$, and (\textbf{c})  $m_z$ in the EP-SHO regime at $\Idc = 2.484 \,\textrm{mA}$. the simulations data show switching from the large-amplitude regime of persistent auto-oscillations to a static state of magnetization. (\textbf{d}) All three components, $m_x$, $m_y$ and $m_z$ are shown over a shorter time scale to illustrate details of the auto-oscillatory dynamics.
\label{fig:Time_domain}}
 \end{figure}
Simulations of the easy-plane spin Hall oscillator (EP-SHO) dynamics reveal bi-stable behavior of the system in the presence of $\Idc$: the system can switch from large amplitude dynamic state to static state  at fixed current above the critical current. Supplementary Fig. \ref{fig:Time_domain} shows example of this behavior for the maximum amplitude dynamic state achieved in simultations at $\Idc = 2.484 \, \mathrm{mA}$. This behavior warrants using time domain analysis of the dynamic state to convert magnetization oscillations to resistance oscillations using Eq. (2) of the main text as opposed to fast Fourier transforms (FFT) as used for the standard spin Hall oscillator (S-SHO) configuration. We include the temporal transient regime in order to capture the shifting frequency modes with increasing current. The system is evolved for $75 \, \mathrm{ns}$.  
These results are shown in main text Fig. 5 (b) and (d).

\subsection{D. S-SHO dynamics visualization.}
\begin{figure}[!tbp]
\center
 \includegraphics[width= \linewidth]{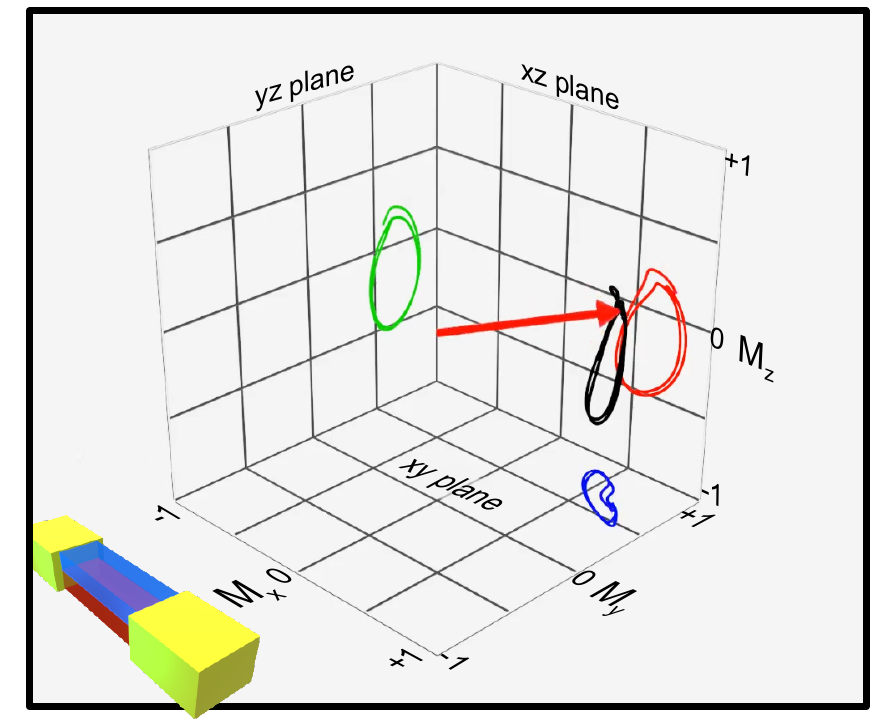}
\caption{\textbf{Micromagnetic simulations of S-SHO auto-oscillations at} \bm{$\Idc = 2.4949\, \mathrm{mA} $}. Average magnetization of the active region traced over 2 periods of auto-oscillations.
\label{fig:S-SHO_snapshot}}
\end{figure}

Supplementary Movie 3 shows dynamics of the average magnetization in the active region for the S-SHO configuration. Supplementary Fig. \ref{fig:S-SHO_snapshot} displays a frame from this Movie showing the path traced by the average magnetization over 2 periods of auto-oscillations. 

\bibliography{refer}